\documentclass[draftclsnofoot,twoside,onecolumn,letter,12pt]{IEEEtran}
\usepackage{graphicx}
\usepackage{adjustbox}
\usepackage{amssymb}
\usepackage{amsmath,cases}
\usepackage{cite}
\usepackage{stfloats}
\usepackage{subfigure}
\usepackage{epstopdf}
\usepackage{mdwlist}
\usepackage{threeparttable}
\usepackage[all,2cell]{xy} \UseAllTwocells
\usepackage{booktabs}
\usepackage{fancyvrb}
\usepackage{rotating}
\usepackage{rotfloat}
\usepackage{pifont}
\usepackage{verbatim}

\usepackage{bm}

\DeclareMathAlphabet{\mathsfbr}{OT1}{cmss}{m}{n}%for math sans serif (cmss)
\SetMathAlphabet{\mathsfbr}{bold}{OT1}{cmss}{bx}{n}%for math sans serif (cmss)
\DeclareRobustCommand{\msf}[1]{%
  \ifcat\noexpand#1\relax\msfgreek{#1}\else\mathsfbr{#1}\fi%for math sans serif (cmss)
}
\makeatletter
\newcommand{\msfgreek}[1]{\csname s\expandafter\@gobble\string#1\endcsname}
\makeatother

\DeclareFontEncoding{LGR}{}{} % or load \usepackage{textgreek}
\DeclareSymbolFont{sfgreek}{LGR}{cmss}{m}{n}
\SetSymbolFont{sfgreek}{bold}{LGR}{cmss}{bx}{n}
\DeclareMathSymbol{\salpha}{\mathord}{sfgreek}{`a}
\DeclareMathSymbol{\sbeta}{\mathord}{sfgreek}{`b}
\DeclareMathSymbol{\sgamma}{\mathord}{sfgreek}{`g}
\DeclareMathSymbol{\sdelta}{\mathord}{sfgreek}{`d}
\DeclareMathSymbol{\sepsilon}{\mathord}{sfgreek}{`e}
\DeclareMathSymbol{\szeta}{\mathord}{sfgreek}{`z}
\DeclareMathSymbol{\seta}{\mathord}{sfgreek}{`h}
\DeclareMathSymbol{\stheta}{\mathord}{sfgreek}{`j}
\DeclareMathSymbol{\siota}{\mathord}{sfgreek}{`i}
\DeclareMathSymbol{\skappa}{\mathord}{sfgreek}{`k}
\DeclareMathSymbol{\slambda}{\mathord}{sfgreek}{`l}
\DeclareMathSymbol{\smu}{\mathord}{sfgreek}{`m}
\DeclareMathSymbol{\snu}{\mathord}{sfgreek}{`n}
\DeclareMathSymbol{\sxi}{\mathord}{sfgreek}{`x}
\DeclareMathSymbol{\somicron}{\mathord}{sfgreek}{`o}
\DeclareMathSymbol{\spi}{\mathord}{sfgreek}{`p}
\DeclareMathSymbol{\srho}{\mathord}{sfgreek}{`r}
\DeclareMathSymbol{\ssigma}{\mathord}{sfgreek}{`s}
\DeclareMathSymbol{\stau}{\mathord}{sfgreek}{`t}
\DeclareMathSymbol{\supsilon}{\mathord}{sfgreek}{`u}
\DeclareMathSymbol{\sphi}{\mathord}{sfgreek}{`f}
\DeclareMathSymbol{\schi}{\mathord}{sfgreek}{`q}
\DeclareMathSymbol{\spsi}{\mathord}{sfgreek}{`y}
\DeclareMathSymbol{\somega}{\mathord}{sfgreek}{`w}

\DeclareMathSymbol{\svarsigma}{\mathord}{sfgreek}{`c}

\DeclareMathSymbol{\sGamma}{\mathalpha}{sfgreek}{`G}
\DeclareMathSymbol{\sDelta}{\mathalpha}{sfgreek}{`D}
\DeclareMathSymbol{\sTheta}{\mathalpha}{sfgreek}{`J}
\DeclareMathSymbol{\sLambda}{\mathalpha}{sfgreek}{`L}
\DeclareMathSymbol{\sXi}{\mathalpha}{sfgreek}{`X}
\DeclareMathSymbol{\sPi}{\mathalpha}{sfgreek}{`P}
\DeclareMathSymbol{\sSigma}{\mathalpha}{sfgreek}{`S}
\DeclareMathSymbol{\sUpsilon}{\mathalpha}{sfgreek}{`U}
\DeclareMathSymbol{\sPhi}{\mathalpha}{sfgreek}{`F}
\DeclareMathSymbol{\sPsi}{\mathalpha}{sfgreek}{`Y}
\DeclareMathSymbol{\sOmega}{\mathalpha}{sfgreek}{`W}

\DeclareRobustCommand{\mcal}[1]{%
  \ifcat\noexpand#1\relax\mathnormal{#1}\else\cal{#1}\fi
}
\DeclareRobustCommand{\BM}[1]{%
  \ifcat\noexpand#1\relax\bm{\boldUppercaseItalicGreek{#1}}\else\bm{#1}\fi
}
\makeatletter
\newcommand{\boldUppercaseItalicGreek}[1]{\csname var\expandafter\@gobble\string#1\endcsname}
\makeatother
\newcommand{\rv}[1]{\msf{#1}}

\let\geq\geqslant

\let\leq\leqslant

\DeclareMathAlphabet{\mathpzc}{OT1}{pzc}{m}{it}

\newcommand{\BB}[1]{\pmb{#1}}
\newcommand{\pDefine}[1]{\mathpzc{#1}}

\newcommand{\oSeq}{\BB{\pDefine{O}}}

\newcommand{\pSeq}{\BB{\pDefine{P}}}
\newcommand{\pC}{\pDefine{c}}
\newcommand{\pM}{\pDefine{m}}
\newcommand{\pN}{\pDefine{n}}
\newcommand{\pP}{\pDefine{p}}
\newcommand{\pQ}{\pDefine{q}}
\newcommand{\pK}{\pDefine{k}}
\newcommand{\pA}[1]{\pDefine{a}_{#1}}
\newcommand{\pB}[1]{\pDefine{b}_{#1}}
\newcommand{\pS}[1]{\pDefine{A}_{#1}}
\newcommand{\pT}[1]{\pDefine{B}_{#1}}
\newcommand{\aV}{\BB{\pDefine{a}}}
\newcommand{\bV}{\BB{\pDefine{b}}}
\newcommand{\sV}{\BB{\pDefine{A}}}
\newcommand{\tV}{\BB{\pDefine{B}}}

\newcommand{\FoxD}[1]{\mathscr{H}\left(#1\right)}
\newcommand{\Fox}[5]{H^{#1,#2}_{#3,#4}\left(#5\right)}

\newcommand{\FoxH}[7]{H^{#1,#2}_{#3,#4}
                        \left[
                            {#5}
                            \left|
                            \begin{array}{c}
                                {#6}
                                \\
                                {#7}
                            \end{array}
                            \right.
                        \right]}

\newcommand{\FoxRV}[2]{\mathscr{H}\left(#1,#2\right)}

\newcommand{\FoxV}[5]{\mathscr{H}^{#1,#2}_{#3,#4}\left(#5\right)}

\newcommand{\stdOP}[2]{#1 \boxdot #2}
\newcommand{\canOP}[2]{#1 \boxplus #2}
\newcommand{\eOP}[3]{\Bra{ #1,#2,#3}}

\newcommand{\FoxHT}[5]{\mathbbmss{H}^{#1}_{#2}\left\{#4;#3\right\}\left(#5\right)}

\newcommand{\FontDef}[1]{\text{#1}}

\newcommand{\pSeqcdf}{\pSeq_\FontDef{cdf}}
\newcommand{\pSeqexp}{\pSeq_\FontDef{exp}}

\usepackage{color}
\usepackage{xcolor}

\usepackage{color}
\usepackage{xcolor}
\usepackage{dsfont}
\usepackage{bbm}
\usepackage[mathscr]{euscript}
\usepackage{braket}
\usepackage{mathtools}

\DeclareMathAlphabet{\mathpzc}{OT1}{pzc}{m}{it}

\DeclareMathAlphabet{\mathitsf}{OML}{cmbr}{m}{it}
\definecolor{CCTLABgreen}{RGB}{0,128,0}

\AtBeginDocument{}

\AtBeginDocument{}

\newtheorem{definition}{Definition}
\newtheorem{theorem}{Theorem}

\newtheorem{corollary}{Corollary}
\newtheorem{proposition}{Proposition}

\newtheorem{remark}{Remark}

\newtheorem{keynote}{Keynote}

\newtheorem{problem}{Problem}

\DeclareMathOperator{\E}{\mathds{E}}

\DeclareMathOperator{\prob}{\mathds{P}}

\DeclareMathOperator{\sign}{\mathrm{sign}}
\DeclareMathOperator{\sgn}{\mathrm{sgn}}

\newcommand{\N}{\mathbbmss{N}}
\newcommand{\R}{\mathbbmss{R}}
\newcommand{\C}{\mathbbmss{C}}

\newcommand{\B}[1]{\mathbf{#1}}

\providecommand\BB{}
\renewcommand{\BB}[1]{\pmb{#1}}

\newcommand{\EX}[1]{\E\left\{{#1}\right\}}

\newcommand{\PDF}[2]{p_{{#1}}\left({#2}\right)}
\newcommand{\CDF}[2]{F_{{#1}}\left({#2}\right)}
\newcommand{\MGF}[2]{\phi_{#1}\left({#2}\right)}

\newcommand{\Prob}[1]{\prob\left\{{#1}\right\}}
\newcommand{\CF}[2]{\phi_{#1}\left({#2}\right)}

\newcommand{\deq}{\mathop{=}\limits^{\text{d}}}

\newcommand{\sNorm}[1]{\left|{#1}\right|}

\newcommand{\argmax}{\mathop{\mathrm{arg\,max}}}

\newcommand{\IndF}[2]{\mathbbmss{1}_{#1}\left(#2\right)}

\newcommand{\bTRe}{\begin{dingautolist}{182}}
\newcommand{\eTRe}{\end{dingautolist}}

\newcommand{\mcsnr}{\mathrm{SNR}}

\newcommand{\rp}[2]{\msf{#1}\left({#2}\right)}

\newcommand{\mean}[1]{\left<{#1}\right>}

\newcommand{\GF}[1]{\Gamma\left(#1\right)}

\newcommand{\LT}[2]{\mathbbmss{L}\left\{{#1}\right\}\left({#2}\right)}

\newcommand{\GMP}[1]{\mathcal{P}\left({#1}\right)}

\newcommand{\STalpha}{{\alpha_\mathrm{st}}}
\newcommand{\STbeta}{{\beta_\mathrm{st}}}

\newcommand{\EKalpha}{{\alpha_\mathrm{ek}}}
\newcommand{\EKbeta}{{\beta_\mathrm{ek}}}

\newcommand{\Stable}[4]{\mathscr{S}\left({#1},{#2},{#3},{#4}\right)}

\newcommand{\Salpha}{\alpha}%{{\alpha_\mathrm{s}}}
\newcommand{\Sbeta}{\beta}%{{\beta_\mathrm{s}}}
\newcommand{\Sgamma}{\gamma}%{{\gamma_\mathrm{s}}}
\newcommand{\Smu}{\mu}%{{\mu_\mathrm{s}}}

\newcommand{\FoxSV}[5]{\bar{\mathscr{H}}^{#1,#2}_{#3,#4}\left(#5\right)}
\newcommand{\FoxSD}[1]{\bar{\mathscr{H}}\left(#1\right)}

\begin{document}

\title{
        Molecular Communication in $H$-Diffusion
	%\\[0.5cm]
	}
\author{
	\vspace{1cm}
        Dung Phuong Trinh, %~\IEEEmembership{Student Member,~IEEE},
        Youngmin~Jeong, %~\IEEEmembership{Member,~IEEE}, \\
        Hyundong~Shin, %~\IEEEmembership{Senior Member,~IEEE},
        and
        Moe~Z.~Win %,~\IEEEmembership{Fellow,~IEEE}\\

\thanks{
        D.~P.~Trinh, Y.~Jeong, and H.~Shin are with the Department of Electronic Engineering,
        Kyung Hee University,
        1732 Deogyeong-daero, Giheung-gu,
        Yongin-si, Gyeonggi-do 17104 Korea
        (e-mail: \{dungtp, yjeong, hshin\}@khu.ac.kr).
}
\thanks{M.~Z.~Win is with
            the Laboratory for Information and Decision Systems (LIDS),
            Massachusetts Institute of Technology,
            77 Massachusetts Avenue,
            Cambridge, MA 02139 USA.
            (e-mail: moewin@mit.edu).
}
%\thanks{
%        D.~P.~Trinh and Y.~Jeong are co-first authors.
%}
}

\maketitle %\thispagestyle{empty}

\begin{abstract}

The random propagation of molecules in a fluid medium is characterized by the spontaneous diffusion law as well as the interaction between the environment and molecules. In this paper, we embody the anomalous diffusion theory for modeling and analysis in molecular communication. We employ \emph{H-diffusion} to model a non-Fickian behavior of molecules in diffusive channels. $H$-diffusion enables us to model anomalous diffusion as the subordinate relationship between self-similar parent and directing processes and their corresponding probability density functions with two $H$-variates in a unified fashion.  
In addition, we introduce standard $H$-diffusion to make a bridge of normal diffusion across well-known anomalous diffusions such as space-time fractional diffusion, Erd\'elyi-Kober fractional diffusion, grey Brownian motion, fractional Brownian motion, and Brownian motion.  We then characterize the statistical properties of uncertainty of the random propagation time of a molecule governed by $H$-diffusion laws by introducing a general class of molecular noise---called $H$-noise. Since $H$-noise can be an algebraic tailed process, we provide a concept of $H$-noise power using finite logarithm moments based on zero-order statistics. Finally, we develop a unifying framework for error probability analysis in a timing-based molecular communication system with a concept of signal-to-noise power ratio. 
%defined by the time resource and $H$-noise.

\end{abstract}


\newpage

\section{Introduction}		\label{sec:1}

Nanotechnology has been highlighted for various applications such as medical systems, healthcare systems, nano-material, nano-machinary, nanoscale communication networks, and molecular communication systems. %The possibility of nanotechnology was first stated by physicist Richard Feynman at an American Physical Society meeting in 1959, with his lecture  \emph{``There's Plenty of Room at the Bottom''} \cite{Fey:60:Caltech}. 
%The concept of nanoscale communication has been recently distinguished more precisely with the development of essential elements of nanoscale communication systems. For example, the IEEE P1906.1 developed a standard for nanoscale and molecular communication to expand and/or overcome the capabilities of nanomachines  \cite{1906:1:16}. 
%
In particular, molecular communication is an emerging technology for communication between nanomachines where information is conveyed by means of molecules \cite{SEA:12:IT,NEH:13:Book, FMGCEG:19:MBSC}. In \emph{passive transport} molecular communication, the random propagation of molecules in a fluid medium such as air, water, or blood vessels in human tissue can be solely determined by the law of diffusion or can be subjected to unpredictable turbulence caused by the environment.\footnote{According to the type of molecule propagation, the mechanisms of molecular communication can be classified into \emph{active transport} and \emph{passive transport}. In active transport molecular communication, molecules propagate through predefined pathways such as a molecular motor, while in passive transport molecular communication, molecules are released through spontaneous diffusion.} Therefore, it is crucial to consider for the spontaneous diffusion law as well as non-predictable turbulence in the fluid medium in characterizing molecular communication.
%\cite{SEA:12:IT, KC:13:JSAC,PA:14:COM,FYECG:16:CSTO,NEH:13:Book}

Brownian motion describes an ideal diffusion environment where the movement of molecules in the fluid medium is induced by their collisions. This diffusion can be modeled by Fick's laws where the homogeneous diffusion coefficient can be applied both in space and time, assuming that each molecule propagates independently. Due to the mathematical convenience---i.e., the time evolution of the probability density function (PDF) associated with the position of the molecule is normally distributed with zero mean---rather than the accuracy of diffusion models, Brownian motion has been widely used to model a molecular communication channel (see, e.g., \cite{SEA:12:IT,KA:13:JSAC,JAJSS:16:COM,LZMY:17:CL,LHLCJ:17:MNL,CLY:18:NB} and references therein). However, the extraordinary diffusion phenomenon, called \emph{anomalous diffusion} or non-Fickian diffusion, was discovered in crowded, heterogeneous, or complex structure systems, (e.g., the particles in heterogeneous porous media \cite{KB:88:PF, Plo:14:JAM}, in cytoplasm \cite{RVD:13:BJ}, and in the rotating flow \cite{SWS:93:PRL}, magnetic resonance in excised human tissue \cite{OBS:06:JMR}, telomeres in nucleus of mammalian cells \cite{BIK:09:PRL}) where the anomalous diffusion process does not obey the linear relationship between mean square displacement and time, in contrast to the ideal diffusion process \cite{MK:00:PR, MJCB:14:PCCP, ZDK:15:RMP,MMM:16:NB}. This highly motivates the use of anomalous diffusion in molecular communication for a wide range of applications \cite{CTJS:15:CL, TJS:19:ACCESS, TJSW:18:COM}.\footnote{Experimental data obtained using molecular communication platforms showed that the end-to-end molecular communication channel has a nonlinearity, unlike many previously developed molecular communication channel models \cite{FKEC:14:JSAC}. } In \cite{CTJS:15:CL}, the one-dimensional anomalous diffusion propagation was considered and the error rate was analyzed for timing and amplitude modulations. This work further extended to a connectivity problem with a random time constraint in a one-dimensional nanonetwork, where the random locations of molecules at the initial time are modeled by poisson point process \cite{TJS:19:ACCESS}. The Cox process has been considered in \cite{TJSW:18:COM} to capture the dynamic variation of the molecule concentration arising from the mobility of anomalously diffusive molecules, and the spatial ordering of the molecular communication performance has been characterized in terms of the error rate in the presence of interfering molecules. However, there is no comprehensive study on the modeling of anomalous diffusion channels in the context of molecular communication.
%\cite{SEA:12:IT,KC:13:JSAC,KA:13:JSAC,PA:14:COM,SML:14:WCOML,FRV:15:NB,LHL:15:COM,JAJSS:16:COM,GA:16:COM,LZMY:17:CL,LHLCJ:17:MNL,CLY:18:NB}

%Discovered by Lewis F. Richardson in 1926 from his large volume of experimental data \cite{Ric:26:PRSA}, the anomalous diffusion theory has been studied to understand its underlying mechanism. 

Various anomalous diffusion processes typically can be modeled numerous ways including continuous random walk (CTRW), generalized diffusion equation, generalized master equation, fractional Brownian motion, and fractional kinetic equation (fractional diffusion equation) \cite{MK:00:PR,WR:83:ACP,Sch:93:PRE,Tun:74:SP}.\footnote{The master equation for Brownian motion is the standard linear diffusion equation. Its fundamental solution is coined with the Gaussian density function, where the spatial variance increases linearly in time.}
In particular, the CTRW simply describes diffusion of molecules in the medium with arbitrary distributions of jump lengths and waiting times.\footnote{The CTRW can be considered as diffusion governed by a space-time fractional diffusion equation, where probabilities for jump length and waiting time behave asymptotically like powers with negative exponents related to the orders of the fractional derivatives, which is called \emph{parametric subordination} \cite{GMV:07:CSF,GVM:06:MS}.} In addition, the combination of a stochastic operational time---a \emph{directing process}---and the \emph{self-similar parent process} is equivalent to the subordination integral mechanism for the product of two random variables in the context of subordinated processes \cite{MPG:03:FCAA,PP:16:FCAA,MP:08:JPA,PMM:13:PTRSA}.\footnote{A natural way to derive subordination formulas for the fractional diffusion processes is called the \emph{stochastic method}  \cite{BM:01:FCAA,MBS:02:PRE,CKN:17:SPL}, in contrast to the subordination integral mechanism method.} This subordination law generates the solution of the fractional diffusion equation in purely analytical ways using the machinery offered by convolution properties of the Mellin transform.\footnote{The Mellin and convolution operators are essential to provide a systematic language in dealing with communication performance by averaging of nonnegative random variables in wireless communication systems \cite{JCSW:13:JSAC,JQKS:14:WCOM,JSW:15:IT,TJS:17:ACCESS} using a general integral transform---called an $H$-transform.}

In this paper, we embody anomalous diffusion according to the self-similar processes. To this end, we introduce the $H$-process by capturing the concept of $H$-variate that is the versatile family of statistical distributions \cite{JSW:15:IT}. Therefore, the $H$-process is well diversified including various typical stochastic processes as special cases. Furthermore, the modeling and analysis framework based on $H$-theory falls into the $H$-transform framework \cite{JSW:15:IT, TJSW:18:COM}; hence, the framework serves as a systematic method in a unified fashion to model the anomalous diffusion channel and analyze the molecular communication system by the virtue of Mellin and convolution operators of two $H$-functions, covering various typical diffusion models. The main contributions of this paper can be summarized as follows:

\begin{itemize}

\item

By introducing a new class of stochastic self-similar processes, namely an $H$-process and a symmetric $H$-process (see Definition~\ref{def:Hprocess} and Definition~\ref{def:SHprocess}), we show that the parent-directing subordination process---which consists of the parent $H$-process and directing $H$-process---is again an $H$-process (see Theorem~\ref{thm:subordination}). Using these two self-similar $H$-processes, we introduce \emph{$H$-diffusion} (see Definition~\ref{def:Hdiffusion}), which can encompass most typical anomalous diffusion types such as space-time fractional diffusion (ST-FD), space fractional diffusion (S-FD), time fractional diffusion (T-FD), Erd\'elyi-Kober fractional diffusion (EK-FD), and grey Brownian motion (GBM) as its special cases. Specifically, we define standard $H$-diffusion (SHD) which allows to model fractional Brownian motion (FBM) and Brownian motion (BM) as well as typical anomalous types of $H$-diffusion as special cases (see Definition~\ref{def:StdHdiffuion}).

\item 
We present a new class of molecular noise---namely, \emph{$H$-noise}---to develop a unifying framework for characterizing statistical properties of uncertainty of the random propagation time of a molecule (see Definition~\ref{def:Hnoise}). We derive the first passage time (FPT) of an anomalous diffusive molecule in terms of $H$-variate (see Theorem~\ref{thm:fpt:Hdiffusion}), which acts as the unique source of uncertainty in diffusive molecular communication. The $H$-noise is well-diversified to various types of molecular noise models such as L{\'e}vy distribution noise for various diffusion scenarios. Then, we put forth an $H$-noise tail (see Theorem~\ref{thm:Hnoise:tails}) and finite logarithm moments of $H$-noise (see Theorem~\ref{thm:LM:Hnoise}) to describe algebraic-tailed and heavy-tailed distribution properties of the $H$-noise. We quantify the geometric power of $H$-noise---namely, \emph{$H$-noise power} coined with the fractional lower-order statistics and zero-order statistics (see Corollary~\ref{cor:gmp:Hnoise} and Remark~\ref{rem:GPF}). 

\item
We characterize the effect of anomalous diffusion on the error probability in timing-based molecular communication. We first develop the unifying error probability analysis for multiple number of transmitted molecules with the $M$-ary modulation technique (see Theorem~\ref{thm:SEP:Hdiffusion}). The first arrival detection method is adopted to carry out the simplified decoding process at a receive nanomachine (RN) using the statistic of the first arrival molecule among multiple released molecules at a transmit nanomachine (TN) (see Proposition~\ref{pro:pdf:fa}). For SHD, we define a signal-to-noise ratio (SNR) based on the geometric power of $H$-noise (see Definition~\ref{def:snr}) and characterize the high-SNR behavior of the error probability in terms of the high-SNR slope and high-SNR power offset for anomalous diffusion-based molecular communication (see Corollary~\ref{cor:highsnr}). 

\end{itemize}

Throughout the paper, we shall adopt the notation that random variables are displayed in sans serif, upright font; their realizations in serif, italic font. For example, a random variable and its realization are denoted by $\rv{x}$ and $x$, respectively. The basic operations on the $H$-function can be found in Table~\ref{table:operation} (see also \cite{JSW:15:IT}).  We relegate the glossary of notations and symbols used in the paper to Appendix~\ref{sec:appendix:NS}.  In Appendix~\ref{sec:appendix:SF}, we briefly introduce the special functions which are frequently used in the context of diffusion theory, fractional diffusion theory, and molecular communication. $H$-representation for stable distributions is provided in Appendix~\ref{sec:appendix:Stable}, which is required to develop a framework for modeling and analysis in  molecular communication.

\begin{table}%[t]	
\caption{Operations on the order and parameter sequences of Fox's $H$-function}
\centering
\label{table:operation}

\begin{threeparttable}

%\begin{ruledtabular}
\begin{tabular}{lllll}
%\toprule
\addlinespace
\midrule
\midrule

Operation &
Symbol &
Order or parameter sequence
\\
%\midrule
\midrule
\addlinespace

Scaling &
$\pSeq\ket{\alpha}$ &
$\left(
	\frac{\pK}{\alpha},
	\frac{\pC}{\alpha},
	\aV,
	\bV,
	\sV,
	\tV
\right)$
\\[-0.2cm]
\addlinespace

Conjugate &
$\Bra{\gamma}\pSeq$ &
$\left(
	\frac{\pK}{\pC^{\gamma}},
	\pC,
	\aV+\gamma\sV,
	\bV+\gamma\tV,
	\sV,
	\tV
\right)$
\\[-0.2cm]
\addlinespace

Elementary &
$\eOP{\alpha}{\beta}{\gamma}\pSeq$ &
$\left(
	\frac{\pK}{\left(\alpha\pC\right)^{\beta\gamma}},
	\left(\alpha\pC\right)^{\beta},
	\aV+\beta\gamma\sV,
	\bV+\beta\gamma\tV,
	\beta\sV,
	\beta\tV
\right)$
\\[-0.2cm]
\addlinespace

Inverse &
$\oSeq^{-1}$ &
$\left(
	\pN,
	\pM,
	\pQ,
	\pP
\right)$
\\[-0.2cm]
\addlinespace

&
$\pSeq^{-1}$ &
$\left(
	\pK,
	\frac{1}{\pC},
	\B{1}_\pQ - \bV,
	\B{1}_\pP - \aV,
	\tV,
	\sV
\right)$
\\[-0.2cm]
\addlinespace

Mellin &
$\stdOP{\oSeq_1}{\oSeq_2}$ &
$\left(
	\pM_1+\pN_2,
	\pN_1+\pM_2,
	\pP_1+\pQ_2,
	\pQ_1+\pP_2
\right)$
\\[-0.2cm]
\addlinespace

&
$\stdOP{\pSeq_1}{\pSeq_2}$ &
$\left(
 	\frac{\pK_1\pK_2}{\pC_2},
	\frac{\pC_1}{\pC_2},
	\aV,
	\bV,
	\sV,
	\tV
\right)$
\\[-0.2cm]
\addlinespace

&
&

$
\begin{cases}
\aV=
		\left(
			\dot{\aV_{1}},
			\B{1}_{\pQ_2}-\bV_{2}-\tV_{2},
			\ddot{\aV_{1}}
		\right)
\\
\bV=
		\bigl(
			\dot{\bV_{1}},
			\B{1}_{\pP_2}-\aV_{2}-\sV_{2},
			\ddot{\bV_{1}}
		\bigr)
\\
\sV=
		\left(
			\dot{\sV_{1}},
			\tV_{2},
			\ddot{\sV_{1}}
		\right)
\\
\tV=
		\left(
			\dot{\tV_{1}},
			\sV_{2},
			\ddot{\tV_{1}}
		\right)
\end{cases}
$

\qquad
(or \cite[Eqs.~(30), (31)]{JSW:15:IT})
\\[-0.2cm]
\addlinespace

Convolution &
$\canOP{\oSeq_1}{\oSeq_2}$ &
$\left(
	\pM_1+\pM_2,
	\pN_1+\pN_2,
	\pP_1+\pP_2,
	\pQ_1+\pQ_2
\right)$
\\[-0.2cm]
\addlinespace

&
$\canOP{\pSeq_1}{\pSeq_2}$ &
$\left(
        		\pK_1 \pK_2,
        		\pC_1 \pC_2,
		\aV,
		\bV,
		\sV,
		\tV
\right)$
\\[-0.2cm]
\addlinespace

&
&
$
\begin{cases}
\aV=
		\left(
			\dot{\aV_{1}},
			\aV_{2},
			\ddot{\aV_{1}}
		\right)
\\
\bV=
		\bigl(
			\dot{\bV_{1}},
			\bV_{2},
			\ddot{\bV_{1}}
		\bigr)
\\
\sV=
		\left(
			\dot{\sV_{1}},
			\sV_{2},
			\ddot{\sV_{1}}
		\right)
\\
\tV=
		\left(
			\dot{\tV_{1}},
			\tV_{2},
			\ddot{\tV_{1}}
		\right)
\end{cases}
$
\hspace{1.8cm} (or \cite[Eqs.~(42), (43)]{JSW:15:IT})
\\
\addlinespace
\midrule
\midrule
%\bottomrule
\end{tabular}
%\end{ruledtabular}

\begin{tablenotes}
\item[\hspace{0.4cm}Note)]
$\alpha, \beta \in \R_{++}$, $\gamma \in \C$,
$\ell \in \N$

\end{tablenotes}

\end{threeparttable}

\end{table}

\section{$H$-Diffusion Modeling}		\label{sec:2}

In this section, we begin by reviewing the subordination law with self-similar processes for modeling anomalous diffusion \cite{MP:08:JPA,PMM:13:PTRSA}. Then, we introduce an $H$-diffusion model, which can span a wide range of well-established anomalous diffusion types.

\subsection{$H$-Variables}

We use the representation of Fox's $H$-function defined in \cite{JSW:15:IT} (see also Appendix~\ref{sec:appendix:NS}) for notational simplicity as \cite[Eq.~(245)]{JSW:15:IT}:
\begin{align}
\Fox{\pM}{\pN}{\pP}{\pQ}{x;\pSeq}
&=
%	\begin{cases}
		\pK\FoxH{\pM}{\pN}{\pP}{\pQ}{\pC x}{\left(\aV,\sV\right)}{\left(\bV,\tV\right)} 
	\qquad
	\left(x>0\right),
\end{align}
where
$
\pSeq
=
	\left(\pK,\pC,\aV,\bV,\sV,\tV\right)
$
is the parameter sequence satisfying the necessary conditions \cite[Remark~7]{JSW:15:IT} to be a density function. By convention, letting the \emph{null} sequences be $\pSeq_\emptyset =\left(1,1,\text{--},\text{--},\text{--},\text{--}\right)$ and $\oSeq_\emptyset=\left(0, 0, 0, 0\right)$, we define
\begin{align}
\Fox{0}{0}{0}{0}{x;\pSeq_\emptyset}
=
	\delta\left(x-1\right).
\end{align}

\begin{definition}[$H$-Variable \cite{JSW:15:IT}] \label{def:H}
A nonnegative random variable $\rv{x}$ is said to have an $H$-distribution with the order sequence $\oSeq=\left(\pM,\pN,\pP,\pQ\right)$ and the parameter sequence $\pSeq = \left(\pK, \pC, \aV, \bV, \sV, \tV \right)$, denoted by $\rv{x} \sim \FoxRV{\oSeq}{\pSeq}$ or simply $\rv{x} \sim \FoxV{\pM}{\pN}{\pP}{\pQ}{\pSeq}$, if its PDF is given by
\begin{align} \label{eq:Def:FV}
\PDF{\rv{x}}{x}
&=
        \Fox{\pM}{\pN}{\pP}{\pQ}{x;\pSeq}
        \qquad\quad
        \left(x >0 \right),
\end{align}
with the set of parameters satisfying a distributional structure such that $\PDF{\rv{x}}{x} \geq 0$ for all $x >0$ and $\int_0^\infty \PDF{\rv{x}}{x} dx=1$.
\end{definition}

\begin{definition}[Symmetric $H$-Variable] \label{def:H}
A symmetric random variable $\rv{y}$ is said to have a symmetric $H$-distribution, denoted by $\rv{y} \sim \FoxSV{\pM}{\pN}{\pP}{\pQ}{\pSeq}$, 
if $\sNorm{\rv{y}} \sim \FoxV{\pM}{\pN}{\pP}{\pQ}{\pSeq}$, that is, its PDF is
$
\PDF{\rv{y}}{y}=\frac{1}{2}\Fox{\pM}{\pN}{\pP}{\pQ}{\sNorm{y};\pSeq}
$ for $y \in \R$.
\end{definition}
The cumulative distribution function (CDF), moment generating function (MGF), and moments of the symmetric $H$-variable are given in Table~\ref{table:statistics:FV}. Note that the CDF, MGF, and moments of the $H$-variable can be found in  \cite[Table~IV]{JSW:15:IT}.

\subsection{$H$-Diffusion}

A concept of \emph{diffusion} is originated from modeling the spread of molecular concentration in a medium caused by the random motion of molecules \cite{KS:14:Book}. Some anomalous diffusion processes are well-defined by continuous time random walk \cite{MPG:03:FCAA,GMV:07:CSF,PMM:13:PTRSA} or Gaussian processes with time subordination \cite{MP:08:JPA,PMM:12:IJSA,MM:09:ITSP},  which are also well-defined by a subordinated process with a self-similar parent process and a self-similar directing process. This model enables us to generate the fundamental solution of a given diffusion equation using the Mellin convolution of two independent random variables whose density functions are governed by the subordination law \cite{MP:08:JPA,PMM:13:PTRSA}. In the following theorem, we introduce a versatile family of diffusion established by the subordination law.
%\cite{MPG:03:FCAA,MWW:07:PRE,SX:17:AMP,GMV:07:CSF,PMM:13:PTRSA,MS:08:SPA}

\begin{remark}[Self-Similarity]
%The $H$-process $\left\{\rp{x}{t}; t \geq 0\right\}$ is \emph{self-similar} since $\rp{x}{t} \deq t^{\omega}\rp{x}{1}$ for $t>0$.
A stochastic process $\left\{\rp{s}{t}; t \geq 0\right\}$ is self-similar if, for any $a \in \R_{++}$, there exists $b \in \R_{++}$ such that  \cite{EM:00:JMPB,EM:02:Book}
\begin{align}
\rp{s}{at}
\deq
	b\rp{s}{t}.
\end{align}
If the self-similar process $\left\{\rp{s}{t}\right\}$ is (i) nontrivial,\footnote{A process $\left\{\rp{x}{t}; t\geq 0\right\}$ is a trivial process or a deterministic process if $\rp{x}{t}$ is constant or its distribution $\PDF{\rv{x}\left(t\right)}{x}$ is a $\delta$-distribution for every $t \geq 0$ \cite[Definition~13.6]{Sat:99:Book}.} (ii) stochastically continuous, and (iii) $\rp{s}{0} =0$ almost surely,  
then there exists the self-similarity exponent $\omega \in \R_{++}$ for any $a$ such that $b=a^\omega$, 
and for any $t>0$:
\begin{align}
\rp{s}{t}
\deq
	t^\omega\rp{s}{1}.
\end{align}
 %\footnote{Note that the self-similarity exponent $\omega$ is unique and positive if $\rp{y}{t}$ is nontrivial, self-similar, stochastically continuous at $t=0$, and $\rp{y}{0} = 0$ almost surely \cite{EM:00:JMPB,Lam:62:TAMS}.}
%Note that the $H$-process $\left\{\rp{X}{t}\right\}$ in Definition~\ref{def:Hprocess}

\end{remark}

\begin{definition}[$H$-Process] 	\label{def:Hprocess}
A nonnegative self-similar process $\left\{\rp{x}{t}; t \geq 0\right\}$ is said to be an \emph{$H$-process} with the exponent $\omega \in \R_{++}$, denoted by  $\left\{\rp{x}{t}; t \geq 0\right\} \sim \left\{\FoxV{\pM}{\pN}{\pP}{\pQ}{\pSeq},\omega\right\}$, if $\rp{x}{t} \sim \FoxV{\pM}{\pN}{\pP}{\pQ}{\pSeq\ket{t^{\omega}}}$ for $t >0$. %and $\rp{X}{0} =0$ almost surely. 
%\blue{If the $H$-process including the L\'evy process properties} is increasing as a function of $t$ almost surely, it is called a \emph{$H$-subordinator}.
\end{definition}

\begin{definition}[Symmetric $H$-Process] \label{def:SHprocess}
A symmetric $\left\{\rp{y}{t}; t \geq 0\right\}$ is said to be a \emph{symmetric $H$-process} with the exponent $\omega \in \R_{++}$, denoted by  $\left\{\rp{y}{t}; t \geq 0\right\} \sim \bigl\{\FoxSV{\pM}{\pN}{\pP}{\pQ}{\pSeq},\omega\bigr\}$, if $\rp{y}{t}$ is a self-similar process and $\rp{y}{t} \sim \FoxSV{\pM}{\pN}{\pP}{\pQ}{\pSeq\ket{t^{\omega}}}$ for $t >0$. 
%if $\left\{\sNorm{\rp{Y}{t}}; t \geq 0\right\} \sim \left\{\FoxV{\pM}{\pN}{\pP}{\pQ}{\pSeq;\omega}\right\}$. 
\end{definition}

\begin{table}
\caption{CDF, MGF, and moments of symmetric $H$-variable $\rv{y} \sim \FoxSV{\pM}{\pN}{\pP}{\pQ}{\pSeq}$}\centering%
%$\sNorm{Y} \sim \FoxV{\pM}{\pN}{\pP}{\pQ}{\pSet}$}\centering
\label{table:statistics:FV}

%\begin{ruledtabular}
\begin{threeparttable}

\begin{tabular}{lllll}
%\toprule
\midrule
\midrule
Statistics &
Symbol &
Function representation &
Parameter sequence
\\
%\midrule
\midrule
\addlinespace

CDF &

$\CDF{\rv{y}}{y}$ &

$
	\tfrac{1}{2}+\sgn\left(x\right)\Fox{\pM}{\pN+1}{\pP+1}{\pQ+1}{\sNorm{x};
	\canOP{\pSeqcdf}{\Bra{1}\pSeq}}
$ &
$
\pSeqcdf
=
        \left(
	        	1,
	        	1,
	        	\left(1,\text{--}\right),
	        	\left(\text{--},0\right),
	        	\left(1,\text{--}\right),
	        	\left(\text{--},1\right)
	\right)
$
\\
\addlinespace

MGF &

$\MGF{\rv{y}}{s}$  &
$
	\tfrac{1}{2}\Fox{\pN+1}{\pM}{\pQ}{\pP+1}{s;
	\stdOP{\pSeqexp}{\pSeq}}+\tfrac{1}{2}\Fox{\pN+1}{\pM}{\pQ}{\pP+1}{-s;
	\stdOP{\pSeqexp}{\pSeq}}
$ &
$
\pSeqexp
=
        \left(
	        	1,
	        	1,
	        	\text{--},
	        	\left(0,\text{--}\right),
 	       	\text{--},
	        	\left(1,\text{--}\right)
	\right)
$
\\
\addlinespace

Moment &
$\mean{\rv{y}^\ell}$ &
$
\begin{cases}
	\frac{\pK}{\pC^{\ell+1}}
	\frac{
		\prod_{j=1}^{\pM}\GF{\pB{j}+\left(\ell+1\right)\pT{j}}
		\prod_{j=1}^{\pN}\GF{1-\pA{j}-\left(\ell+1\right)\pS{j}}
	}{
		\prod_{j=\pN+1}^{\pP}\GF{\pA{j}+\left(\ell+1\right)\pS{j}}
		\prod_{j=\pM+1}^{\pQ}\GF{1-\pB{j}-\left(\ell+1\right)\pT{j}}
	},\\
0, \text{if $\ell$ is odd number}
\end{cases}
$ &
$
-
$
\\
\addlinespace
\midrule
\midrule
\end{tabular}
%\end{ruledtabular}
\end{threeparttable}

\end{table}

\begin{theorem}[$H$-Subordination]	\label{thm:subordination}
Let $\left\{\rp{p}{t}; t \geq 0\right\} \sim \left\{\FoxV{\pM_1}{\pN_1}{\pP_1}{\pQ_1}{\pSeq_1},\omega_1\right\}$ be a parent $H$-process
independent of a directing $H$-process $\left\{\rp{d}{t}; t \geq 0\right\} \sim \left\{\FoxV{\pM_2}{\pN_2}{\pP_2}{\pQ_2}{\pSeq_2},\omega_2\right\}$. %nondecreasing almost surely ($H$-subordinator). 
Then, a parent-directing 
subordinated process $\left\{\rp{x}{t}=\rp{p}{\rp{d}{t}}; t \geq 0\right\}$ is again an $H$-process with the self-similar exponent $\omega_1\omega_2$, that is,
\begin{align} \label{eq:gf:hv}
\left\{\rp{x}{t};t \geq 0\right\}
	\sim
	\bigl\{\FoxV{\pM_1+\pM_2}{\pN_1+\pN_2}{\pP_1+\pP_2}{\pQ_1+\pQ_2}
	{
	\pSeq_{\left(\omega_1\right)}},  
	\omega_1 \omega_2 \bigr\}
\end{align}
with  
\begin{align} \label{eq:thm:1}
\pSeq_{\left(\omega_1\right)}
	&=
		\canOP{\pSeq_1}{\eOP{1}{\omega_1}{1/\omega_1-1}\pSeq_2}
\end{align}
where $\eOP{\alpha}{\beta}{\gamma}$, and $\canOP{}{}$ are the elementary and convolution operations on the parameter sequence \cite[Table~III]{JSW:15:IT}.
\begin{proof}
%Since the parent process $\left\{\rp{p}{t}\right\}$ is a L\'evy process and the directing process $\left\{\rp{d}{t}\right\}$ is a subordinator, the subordinated process $\left\{\rp{x}{t}\right\}$ is again a  L\'evy process \cite{App:09:Book}.
%In addition, it follows from the self-similarity of $H$-processes that
Since the parent process $\left\{\rp{p}{t}\right\}$ and the directing process $\left\{\rp{d}{t}\right\}$ are self-similar processes, the subordinated process $\left\{\rp{x}{t}\right\}$ is again a self-similar process as follows:
\begin{align} \label{eq:thm:pf:1}
	\rp{x}{t}
	&\deq
	\rp{p}{t^{\omega_2}\rp{d}{1}}
	\nonumber \\
	&
	\deq
	t^{\omega_1\omega_2}
	\rp{p}{\rp{d}{1}}
	\nonumber \\
	&
	=
	t^{\omega_1\omega_2}
	\rp{p}{1}
	\rp{d}{1}^{\omega_1},
\end{align}
where the last equality follows from the subordination formula
\begin{align}	\label{eq:thm:pf:2}
	\PDF{\rp{x}{1}}{x}
	&=
		\int_0^\infty
		\PDF{\rp{p}{\tau}}{x}
		\PDF{\rp{d}{1}}{\tau}d\tau
	\nonumber \\
	&
	=
		\int_0^\infty
		\PDF{\rp{p}{1}}{\frac{x}{\tau^{\omega_1}}}
		\PDF{\rp{d}{1}}{\tau}
		\frac{d\tau}{\tau^{\omega_1}}.
\end{align}
In addition, $\rp{p}{1}\rp{d}{1}^{\omega_1}$ has an $H$-distribution by \cite[Theorem~1]{JSW:15:IT}. Using \eqref{eq:thm:pf:1}, and the fact that $\alpha \rv{w} \sim \FoxV{\pM}{\pN}{\pP}{\pQ}{\pSeq\Ket{\alpha}}$ if $\rv{w} \sim \FoxV{\pM}{\pN}{\pP}{\pQ}{\pSeq}$ for $\alpha >0$, we have 
\begin{align}  
%	\rp{X}{1}^{\omega_1}
%	\rp{Y}{1}
	\rp{x}{t}
	\sim
	\FoxV{\pM_1+\pM_2}{\pN_1+\pN_2}{\pP_1+\pP_2}{\pQ_1+\pQ_2}
	{
	\pSeq_{\left(\omega_1\right)} 
	\Ket{t^{\omega_1\omega_2}}
	}
\end{align}
which completes the proof.
\end{proof}

\end{theorem}

%\begin{definition}[$H$-Diffusion] 		\label{def:Hdiffusion} 
%Let $\left\{\rp{p}{t}; t \geq 0\right\} \sim \bigl\{\FoxSV{\pM_1}{\pN_1}{\pP_1}{\pQ_1}{\pSeq_1},\omega_1\bigr\}$ 
%be a symmetric $H$-process independent of a directing $H$-process $\left\{\rp{d}{t}; t \geq 0\right\} \sim \left\{\FoxV{\pM_2}{\pN_2}{\pP_2}{\pQ_2}{\pSeq_2},\omega_2\right\}$. Then, a subordinated process $\left\{\rp{x}{t}=\rp{p}{\rp{d}{t}}; t \geq 0\right\}$ is said to be \emph{$H$-diffusion}, denoted by 
%$$\left\{\rp{x}{t}; t \geq 0 \right\} \sim \left\{\FoxSV{\pM_1:\pM_2}{\pN_1:\pN_2}{\pP_1:\pP_2}{\pQ_1:\pQ_2}{\pSeq_1,\pSeq_2;\omega_1,\omega_2}\right\}.$$
%\end{definition}

\begin{definition}[$H$-Diffusion] 		\label{def:Hdiffusion} 
Let $\left\{\rp{p}{t}; t \geq 0\right\} \sim \bigl\{\FoxSV{\pM_1}{\pN_1}{\pP_1}{\pQ_1}{\pSeq_1},\omega_1\bigr\}$ 
be a symmetric $H$-process independent of a directing $H$-process $\left\{\rp{d}{t}; t \geq 0\right\} \sim \bigl\{\FoxV{\pM_2}{\pN_2}{\pP_2}{\pQ_2}{\pSeq_2},\omega_2\bigr\}$. Then, a subordinated process $\left\{\rp{x}{t}=\rp{p}{\rp{d}{t}}; t \geq 0\right\}$ is said to be \emph{$H$-diffusion}, denoted by 
$$\left\{\rp{x}{t}; t \geq 0 \right\} \sim \left\{\FoxSV{\pM_1:\pM_2}{\pN_1:\pN_2}{\pP_1:\pP_2}{\pQ_1:\pQ_2}{\pSeq_1,\pSeq_2;\omega_1,\omega_2}\right\}.$$
\end{definition}

\begin{corollary} \label{crl:Hdiffusion}
Using the same argument in Theorem~\ref{thm:subordination}, we can see that $H$-diffusion is again a symmetric $H$-process with the self-similar exponent $\omega_1\omega_2$:
\begin{align} 
%\left\{\sNorm{\rp{X}{t}};t \geq 0\right\}
\left\{\rp{x}{t};t \geq 0\right\}
	\sim
	\bigl\{\FoxSV{\pM_1+\pM_2}{\pN_1+\pN_2}{\pP_1+\pP_2}{\pQ_1+\pQ_2}
	{
	\pSeq_{\left(\omega_1\right)}},  
	\omega_1 \omega_2 \bigr\}
\end{align}
where $\pSeq_{\left(\omega_1\right)}$ is given in \eqref{eq:thm:1}. If a particle is released into a fluid medium governed by this $H$-diffusion process, at time $t=0$ at position $x=0$, then the particle's position at time $t>0$ is the symmetric $H$-variable
\begin{align} \label{eq:gf:hv}
%\sNorm{\rp{X}{t}}
\rp{x}{t}
	&\deq
	t^{\omega_1\omega_2}
	\rp{p}{1}
	\rp{d}{1}^{\omega_1}
%	Z^{\omega_1}\left(1\right)
\\
	&
	\sim
	\FoxSV{\pM_1+\pM_2}{\pN_1+\pN_2}{\pP_1+\pP_2}{\pQ_1+\pQ_2}
	{
	\pSeq_{\left(\omega_1\right)}  
	\Ket{t^{\omega_1 \omega_2}}}
\end{align}
where $\rp{p}{1}  \sim \FoxSV{\pM_1}{\pN_1}{\pP_1}{\pQ_1}{\pSeq_1}$, and $\rp{d}{1}  \sim \FoxV{\pM_2}{\pN_2}{\pP_2}{\pQ_2}{\pSeq_2}$. 
\end{corollary}

%\begin{remark}[Non-Uniqueness  Representation] \label{rmk:non-unique}
%\blue{
%The PDF $\PDF{\rp{x}{1}}{x}$ of $H$-diffusion is equal to the marginal PDF of the subordinated process $\rp{x}{t}$. This stochastic representation is not unique. In the example below, we define the subordinated process $\rp{d}{t}=\rp{x}{\rp{z}{t}}$ where the parent process $\rp{x}{t}$ is Gaussian process with self similar exponent $1/2$ and the directing process $\rp{z}{t}$ is positive random time process with self similar exponent $\alpha$. We have
%\begin{align}
%\rp{d}{t}=\rp{x}{\rp{z}{t}} \deq \sqrt{\rp{z}{t}}\rp{x}{1} \deq \sqrt{\rp{z}{1}} t^{\alpha/2}\rp{x}{1}\deq \sqrt{\rp{z}{1}}\rp{\tilde{x}}{t}=\rp{y}{t}
%\end{align}
%where $\rp{\tilde{x}}{t}$ has self similar exponent $\alpha/2$. It is clear that the marginal PDF of process $\rp{d}{t}$ coincides with the marginal PDF of process $\rp{y}{t}$ and both $\rp{d}{t}$ and $\rp{y}{t}$ are self similar process with self similar exponent $\alpha/2$. However, while the process $\rp{y}{t}$ always has the stationary increments, it is not always true for the process $\rp{d}{t}$ \cite{MTM:08:PA}.}
%\end{remark}

\begin{remark}[Mean-Square Displacement]	%\label{thm:gf:Hdiffusion}
A particle is released into a fluid medium, governed by this $H$-diffusion process, at time $t=0$ at position $x=0$. Thus, the particle's position at time $t>0$ is the symmetric $H$-variable (see Corollary~\ref{crl:Hdiffusion})
\begin{align} \label{eq:gf:hv}
%\sNorm{\rp{X}{t}}
\rp{x}{t}
	&\sim
	\FoxSV{\pM_1+\pM_2}{\pN_1+\pN_2}{\pP_1+\pP_2}{\pQ_1+\pQ_2}
	{
	\pSeq_{\left(\omega_1\right)} 
	\Ket{t^{\omega_1 \omega_2}}}.
\end{align}
Hence, the mean-square displacement (MSD) of the particle is %(see Table~\ref{table:statistics:FV})
\begin{align}  
\bigl<\sNorm{\rp{x}{t}}^2\bigr>
%\mean{\sNorm{\rp{X}{t}}^2}
%\mean{X^2\left(t\right)}
	\propto
	t^{2\omega_1 \omega_2}
\end{align}
where $2\omega_1 \omega_2$ represents the diffusion exponent. %, which
Using this exponent, we 
can classify diffusion into three types: i) subdiffusion for $0 <\omega_1 \omega_2 < 1/2$, ii)
normal diffusion for $\omega_1 \omega_2=1/2$, and iii) superdiffusion for $\omega_1 \omega_2 >1/2$.

\end{remark}

%\begin{remark}
%The $H$-diffusion can be classified into three types of diffusion depending on the diffusion exponent  $\nu = 2\omega_1 \omega_2$: i) subdiffusion for $0<\nu<1$, ii) normal diffusion for $\nu=1$, and iii) superdiffusion for $\nu >1$, respectively  (see also Fig.~\ref{fig:9}).
%\end{remark}

\begin{remark} \label{rmk:limitation}
Since the $H$-distribution can span a wide range of statistical distributions, $H$-diffusion encompasses various types of anomalous diffusion processes as special cases, including L\'evy flight, space-time fractional diffusion, and Erd\'elyi-Kober fractional diffusion (generalized grey Brownian motion). Moreover, the probability density of molecule location $x$ at given time $t$ can be found from Mellin and convolution operations of two $H$-functions \cite{JSW:15:IT}. %Hence, it provides a diffusion model in both unified and generalized ways. 
%\mynote{Limitation}
However, since the $H$-process needs to be a nontrivial process (it is a necessary condition for the existence of self-similarity exponent), $H$-diffusion cannot cover the family of Brownian motions, whose \emph{directing process} is a trivial process.\footnote{By the definition, the family of Brownian motions is $H$-process but cannot be called $H$-diffusion because its directing process is the $\delta$-distribution.  See Corollary~\ref{corl:fbm} and Table~\ref{table:TDM:HD}.} For example, the directing process of FBM and BW has the form of dirac delta function.
\end{remark}

\begin{corollary} [Fractional Brownian Motion] \label{corl:fbm}
Let $\PDF{\rp{d}{t}}{\tau}=\Fox{0}{0}{0}{0}{\frac{\tau}{t};\pSeq_\emptyset}$. Then, the subordinated process $\left\{\rp{x}{t}; t \geq 0\right\}$ becomes the parent $H$-process $\left\{\rp{p}{t}; t \geq 0\right\} $ as
\begin{align} \label{eq:crl:Hps}
\PDF{\rp{x}{t}}{x}
	&=
		\int_0^\infty
		\PDF{\rp{p}{1}}{\frac{x}{\tau^{\omega_1}}}
		\Fox{0}{0}{0}{0}{\frac{\tau}{t};\pSeq_\emptyset}
		\frac{d\tau}{\tau^{\omega_1}}
		\nonumber \\		
	&
	=
		\int_0^\infty
		\PDF{\rp{p}{1}}{\frac{x}{\tau^{\omega_1}}}
		\delta\left(\frac{\tau}{t}-1\right)
		\frac{d\tau}{\tau^{\omega_1}}
\nonumber \\
	&
	=
		\PDF{\rp{p}{t}}{x}.
\end{align}
With the symmetric Gaussian process of the parent $H$-process, that is,
\begin{align}
\left\{\rp{x}{t}; t \geq 0\right\} 
=
\left\{\rp{p}{t}; t \geq 0\right\} 
\sim \bigl\{\FoxSV{1}{0}{0}{1}{\pSeq_1},\omega_1\bigr\} 
\end{align}
where $\pSeq_1=\Bigl(\frac{1}{2\sqrt{\pi}},\frac{1}{2},\text{--},0,\text{--},\frac{1}{2}\Bigr)$, $\rp{x}{t}$ is FBM. Specifically, $\rp{x}{t}$ is called BM when $\omega_1=1/2$.
\end{corollary}

Table~\ref{table:TDM:HD} shows the typical anomalous diffusion models as special cases of $H$-diffusion and fractional Brownian motion.

%\bookmark

\begin{table}[t]
\caption{%Typical Anomalous Diffusion Models as Special Cases of $H$-diffusion 
Typical Anomalous Diffusion Models as Special Cases of $H$-diffusion and Normal Diffusion
} \centering
\label{table:TDM:HD}

\begin{threeparttable}

\begin{adjustbox}{max width=1\textwidth}
\begin{tabular}{lllllll}
%\toprule
\midrule
\midrule
Diffusion &
\multicolumn{2}{l}{Parent $\rp{p}{1}\sim\FoxSD{\oSeq_1,\pSeq_1}$} &
\multicolumn{2}{l}{Directing $\rp{d}{1}\sim\FoxD{\oSeq_2,\pSeq_2}$} &
&

\\[-0.1cm]

$\rp{h}{t}$ &
$\oSeq_1$ &%=\left(\pM_1,\pN_1,\pP_1,\pQ_1\right)$ &
$\pSeq_1$ &%=\left(\pK_1,\pC_1,\aV_1,\bV_1,\sV_1,\tV_1\right)$ &
$\oSeq_2$ &%=\left(\pM_2,\pN_2,\pP_2,\pQ_2\right)$ &
$\pSeq_2$ &%=\left(\pK_2,\pC_2,\aV_2,\bV_2,\sV_2,\tV_2\right)$ &
$\omega_1$ &
$\omega_2$

\\

%\midrule
\midrule

ST-FD &
$\left(1,1,2,2\right)$ &
$\Bigl(\frac{2}{\alpha},1,\left(1-\frac{1}{\alpha},\frac{1}{2}\right),\left(0,\frac{1}{2}\right),\left(\frac{1}{\alpha},\frac{1}{2}\right),\left(1,\frac{1}{2}\right)\Bigr)$ &
$\left(1,0,1,1\right)$ &
$\left(1,1,1-\beta,0,\beta,1\right)$ &
$1/\alpha$ &
$\beta$ 

\\ \midrule

S-FD &
%$\left(1,1,2,2\right)$ &
%$\Bigl(\frac{2}{\STalpha},1,\left(1-\frac{1}{\STalpha},\frac{1}{2}\right),\left(0,\frac{1}{2}\right),\left(\frac{1}{\STalpha},\frac{1}{2}\right),\left(1,\frac{1}{2}\right)\Bigr)$ &
%$\left(0,0,0,0\right)$ &
%$\left(1,1,\text{--},\text{--},\text{--},\text{--}\right)$ &
%$1/\STalpha$ &
%$-$
$\left(1,1,2,2\right)$ &
$\Bigl(\frac{2}{\alpha},1,\left(1-\frac{1}{\alpha},\frac{1}{2}\right),\left(0,\frac{1}{2}\right),\left(\frac{1}{\alpha},\frac{1}{2}\right),\left(1,\frac{1}{2}\right)\Bigr)$ &
$\left(0,1,1,1\right)$ &
$\left(\frac{\cos\left(\frac{\pi\beta}{2}\right)^{\alpha_d}}{\beta},\cos\left(\frac{\pi\beta}{2}\right)^{\beta},1-\frac{1}{\beta},0,\frac{1}{\beta},1\right)$ &
$1/\alpha$ &
$1/\beta$

\\ \midrule

T-FD &
$\left(1,0,1,1\right)$ &
$\Bigl(1,1,\frac{1}{2},0,\frac{1}{2},1\Bigr)$ &
$\left(1,0,1,1\right)$ &
$\left(1,1,1-\beta,0,\beta,1\right)$ &
$1/2$ &
$\beta$

\\ \midrule

EK-FD &
$\left(1,0,0,1\right)$ &
$\Bigl(\frac{1}{2\sqrt{\pi}},\frac{1}{2},\text{--},0,\text{--},\frac{1}{2}\Bigr)$ &
$\left(1,0,1,1\right)$ &
$\left(1,1,1-\beta,0,\beta,1\right)$ &
$1/2$ &
$\alpha$

\\ \midrule

GBM &
$\left(1,0,0,1\right)$ &
$\Bigl(\frac{1}{2\sqrt{\pi}},\frac{1}{2},\text{--},0,\text{--},\frac{1}{2}\Bigr)$ &
$\left(1,0,1,1\right)$ &
$\left(1,1,1-\beta,0,\beta,1\right)$ &
$1/2$ &
$\beta$

\\ 
\midrule

FBM &
$\left(1,0,0,1\right)$ &
$\Bigl(\frac{1}{2\sqrt{\pi}},\frac{1}{2},\text{--},0,\text{--},\frac{1}{2}\Bigr)$ &
$\left(0,0,0,0\right)$ &
$\left(1,1,\text{--},\text{--},\text{--},\text{--}\right)$ &
$\alpha/2$ &
$-$

\\ \midrule

BM &
$\left(1,0,0,1\right)$ &
$\Bigl(\frac{1}{2\sqrt{\pi}},\frac{1}{2},\text{--},0,\text{--},\frac{1}{2}\Bigr)$ &
$\left(0,0,0,0\right)$ &
$\left(1,1,\text{--},\text{--},\text{--},\text{--}\right)$ &
$1/2$ &
$-$

\\
\midrule
\midrule
%\bottomrule
\end{tabular}

\end{adjustbox}

\begin{tablenotes}

\item[\hspace{0.5cm}(Note)]

$\alpha \in \left(0,2\right]$, $\beta \in \left(0,1\right)$,
$\omega_1,\omega_2\in \R_{++}$

\end{tablenotes}

\end{threeparttable}

\end{table}

%%%%%%%%%%%%%%%%%%%%%%%%%%%%%%%%%%%%%%%%%%

%\mynote{Motivation}
\subsection{Standard $H$-diffusion}
In this subsection, we define SHD coined with special cases of the $H$-process, namely, symmetric L\'evy stable, $M$-Wright, and one-sided L\'evy  stable processes \cite{PMM:13:PTRSA,PP:16:FCAA, MTM:08:PA}. Furthermore, SHD allows to model the FBM and BM as special cases, which have a shared boundary with $H$-diffusion. 

\begin{definition}[Standard $H$-Diffusion]		\label{def:StdHdiffuion}
Let
\begin{align}
\left\{\rp{x}{t}; t \geq 0 \right\} \sim \left\{\FoxV{1:1}{1:0}{2:1}{2:1}{\pSeq_{\left(\alpha_1\right)}\Ket{\beta_1},\pSeq_{\left(\alpha_2\right)}\Ket{\beta_2};
\omega_1,\omega_2}\right\}
\end{align}
with the parameter sequences %$\pSeq_1$ and $\pSeq_2$:
\begin{align}
\pSeq_{\left(\alpha_1\right)}
&=
	\Bigl(
		\tfrac{2}{\alpha_1},
		1,
		\bigl(1-\tfrac{1}{\alpha_1},\tfrac{1}{2}\bigr),
		\bigl(0,\tfrac{1}{2}\bigr),
		\bigl(\tfrac{1}{\alpha_1},\tfrac{1}{2}\bigr),
		\bigl(1,\tfrac{1}{2}\bigr)
	\Bigr)
\nonumber \\
\pSeq_{\left(\alpha_2\right)}
&=
	\Bigl(
		1,
		1,
		1-\alpha_2,
		0,
		\alpha_2,
		1
	\Bigr)
\end{align}
for $\alpha_1 \in \left(0,2\right]$, $\alpha_2 \in \left(0,1\right]$, and $\beta_1,\beta_2\in \R_{++}$ be SHD. Then, the molecule's position $\rp{x}{t}$ governed by SHD is the symmetric $H$-variate
\begin{align}
	\rp{x}{t}
	&\sim
		\FoxSV{2}{1}{3}{3}{\hat{\pSeq}_{\left(\omega_1,\alpha_1,\alpha_2\right)}\Ket{\beta_1\beta_2^{\omega_1}t^{\omega_1\omega_2}}}
\end{align}
where the parameter sequence $\hat{\pSeq}_{\left(\omega_1,\alpha_1,\alpha_2\right)}$ is given by
\begin{align}
\hat{\pSeq}_{\left(\omega_1,\alpha_1,\alpha_2\right)}=
\left(
	\tfrac{2}{\alpha_1},
	1,
	\Bigl(1-\tfrac{1}{\alpha_1},1-\omega_1 \alpha_2,\tfrac{1}{2}\Bigr),
	\Bigl(0,1-\omega_1,\tfrac{1}{2}\Bigr),
	\Bigl(\tfrac{1}{\alpha_1},\omega_1\alpha_2,\tfrac{1}{2}\Bigr),
	\Bigl(1,\omega_1,\tfrac{1}{2}\Bigr)
\right).
\end{align}
\end{definition}

%\begin{corollary}[Green Function of Standard $H$-Diffusion]	\label{cor:GF:sHdiffusion}
%The Green function of SHD $\gfunc{x,t}{\omega_1,\omega_2}$ is corresponding to the PDF of the symmetric $H$-variate $\rv{g}_\mathrm{std}$ such that
%\begin{align}
%	\sNorm{\rv{g}_\mathrm{std}}
%	&\sim
%		\FoxV{2}{1}{3}{3}{\pSeq_\mathrm{std}\Ket{\beta_1\beta_2^{\omega_1}t^{\omega_1\omega_2}}}
%\end{align}
%where the parameter sequence $\pSeq_\mathrm{std}$ is
%\begin{align}
%\pSeq_\mathrm{std}
%&=
%	\biggl(
%			\frac{2}{\alpha_1},
%			1,
%			\left(1-\frac{1}{\alpha_1},\frac{1}{2},1-\omega_1 \alpha_2\right),
%			\left(1-\omega_1,0,\frac{1}{2}\right),
%			\left(\frac{1}{\alpha_1},\frac{1}{2},\omega_1\alpha_2\right),
%			\left(\omega_1,1,\frac{1}{2}\right)
%	\biggr).
%\end{align}
%\begin{proof}
%It follows readily from Theorem~\ref{thm:gf:Hdiffusion}.
%\end{proof}
%\end{corollary}
%

\begin{figure}
    \centerline{\includegraphics[width=0.55\textwidth]{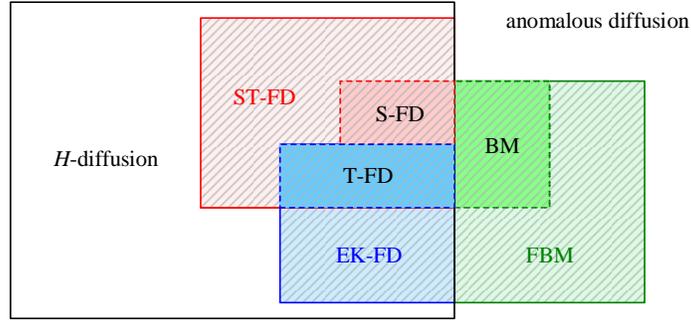}}
    %\vspace{-0.2cm}
    \caption{
        A Venn diagram displays anomalous diffusion sets with their relationship. SHD covers the shaded area.  
    }
    \label{fig:shd}
\end{figure}

\begin{remark}	\label{rem:sHD}
The symmetric $H$-variate $\rp{p}{1}$ in SHD follows a stable distribution with a characteristic exponent $\alpha_1$ and scaling parameter $\beta_1^{\alpha_1}$, i.e., $\rp{p}{1} \sim \Stable{\alpha_1}{0}{\beta_1^{\alpha_1}}{0}$ (see Appendix~\ref{sec:appendix:Stable}), while the $H$-variate $\rp{d}{1}$
is distributed as an $M$-Wright function with the parameter $\alpha_2$ and scaling parameter $\beta_2$ \cite{MMP:10:JDE}. Note that $\rp{d}{1}$ can be obtained from the extremal nonnegative strictly stable random variable $\rv{s}\sim\Stable{\alpha_2}{1}{\cos\left(\pi\alpha_2/2\right)/\beta_2}{0}$ such that
\begin{align}
\rp{d}{1} \sim \rv{s}^{-\alpha_2}.
\end{align}
The stable distribution can explain the heavy-tailed distribution of the stochastic jump process. The $M$-Wright function can be used as a generalization of the Gaussian density for fractional diffusion processes, which plays the key role to describe both slow and fast types of diffusion phenomena in anomalous diffusion \cite{Pag:13:FCAA, MMP:10:JDE, PS:14:CAIM}.\footnote{The nonnegative stable random variable is also known as a possible solution for the waiting time distribution in CTRW. Specifically, the Mittag-Leffler distribution, which has a stretched (natural generalization of) exponential distribution, is used for the waiting time distribution in CTRW as a special case \cite{MRGS:00:PA, Cah:13:CSTM, GKMR:14:Book,Lin:98:SPI}.} Therefore, SHD can well describe the various typical anomalous diffusion models with unit scaling parameters $\beta_1=\beta_2=1$: i) ST-FD with a set of parameters $\left(\alpha_1,\alpha_2,\omega_1,\omega_2\right)=\left(\alpha,\beta,1/\alpha,\beta\right)$; ii) EK-FD with $\left(\alpha_1,\alpha_2,\omega_1,\omega_2\right)=\left(2,\beta,1/2,\alpha\right)$; iii) grey Brownian motion when $\left(\alpha_1,\alpha_2,\omega_1,\omega_2\right)=\left(2,\beta,1/2,\beta\right)$; and iv) standard normal diffusion (Brownian motion) with $\left(\alpha_1,\alpha_2,\omega_1,\omega_2\right)=\left(2,1,1/2,1\right)$. A Venn diagram for anomalous diffusion sets with their relationship is shown in Fig.~\ref{fig:shd}. SHD covers the shaded area.   
\end{remark}

\begin{remark}[Role of Scaling Parameters and Diffusion Coefficient]
Two positive scaling parameters $\beta_1$ and $\beta_2$ in SHD are determined by the diffusion medium. For example, the diffusion equation for ST-FD is well established with the diffusion coefficient $K$, which acts as a scaling factor on the spatial density function. Under our framework, SHD can connect to an equivalent fractional differential diffusion equation form as
\begin{align}   \label{eq:gdiff}
    \frac{\partial ^{\alpha_2}}{\partial t^{\alpha_2}} g\left(x,t\right)
    &=
       \beta_1^{\alpha_1} \beta_2\, \frac{\partial ^{\alpha_1}}{\partial \sNorm{x}^{\alpha_1}} g\left(x,t\right)
\end{align}
where the fundamental solution $g\left(x,t\right)$ can be obtained from the Fourier transform of $\rp{p}{1}$ and the Laplace transform of $\rv{s}$ in Remark~\ref{rem:sHD}. In this case, the diffusion coefficient $K=\beta_1^{\alpha_1} \beta_2$ \cite{VIKH:08:OC}.
\end{remark}

\section{$H$-Noise Modeling}
%\vspace{-0.2cm}
In this section, we characterize the effect of \emph{molecular noise}---influences on the conversation between two nanomachines.  Since the random motion of the molecule emitted from the TN directly effects the uncertainty of absorbing time at the RN, the modeling of molecular noise to unveil the intrinsic characteristic of molecules' movements governed by anomalous diffusion laws is important in designing a molecular communication system.

\subsection{$H$-Noise Model}

Let $\rv{s}$ be the random released time of molecules at the TN. Then, the arrival time $\rv{y}$ of the molecule at the RN located away from the TN is 
\begin{align}	\label{eq:at}
\rv{y}=\rv{s}+\rv{t}
\end{align}
where $\rv{t}$ is an additional random time of the molecule to arrive at the RN. This additional random time $\rv{t}$ evidently plays the role of an \emph{additive} random noise in molecular communication. Since the RN acts as a boundary with a perfect absorbing process, the noise $\rv{t}$ can be thought of as a \emph{FPT} such that the molecule emitted from the TN reaches the boundary for the first time. This random noise has been unveiled as a L\'evy distribution  \cite{NOL:12:CL, KLYFEC:16:JSAC,MMM:16:NB,HAASG:17:CL}, an inverse Gaussian distribution \cite{SEA:12:IT,LZMY:17:CL}, a stable distribution \cite{FMGCEG:19:MBSC}, and (or more generally) $H$-variate \cite{CTJS:15:CL,TJSW:18:COM} for various diffusion scenarios. 

We begin by deriving the FPT of the molecule governed by $H$-diffusion. Then we introduce a general class of molecular noise---namely, \emph{$H$-noise}---to develop a unifying framework for characterizing  statistical properties of uncertainty or distribution of random propagation time.

\begin{theorem}[First Passage Time]		\label{thm:fpt:Hdiffusion}
Let $\left\{\rp{x}{t}; t \geq 0 \right\} \sim \left\{\FoxSV{\pM_1:\pM_2}{\pN_1:\pN_2}{\pP_1:\pP_2}{\pQ_1:\pQ_2}{\pSeq_1,\pSeq_2;\omega_1,\omega_2}\right\}$ and $\rv{t}$ be an FPT, which is defined such that the molecule starting at $x=0$ reaches distance $x=a$, $a\in \R_{+}$ for the first time: 
\begin{align}
	\rv{t}=
	\inf\left\{t: x\left(t\right)=a\right\}.
\end{align}
Given an initial condition $\PDF{\rp{x}{0}}{x}=\delta\left(x\right)$ and a boundary condition $\PDF{\rp{x}{t}}{a}=0$ for the absorbing process, the FPT of the molecule in $H$-diffusion is the $H$-variate:\footnote{The FPT of molecules highly depends on the boundary condition. It is nontrivial to find an accurate statistic of FPT with an arbitrary boundary condition in multi-dimensional space. See \cite{Red:01:Book,TJS:19:ACCESS}.}
\begin{align}	\label{eq:foxv:fpt}
\rv{t}
\sim
	\FoxV{\pN_1+\pN_2}{\pM_1+\pM_2}{\pQ_1+\pQ_2}{\pP_1+\pP_2}
	{
		\pSeq_\rv{t}
		\Ket{\left(a \pC_{\left(\omega_1\right)} \right)^{1/\left(\omega_1\omega_2\right)}}
	}
\end{align}
where the parameter sequence $\pSeq_\rv{t}$ is given by%\footnote{We denote the parameter sequence $\pSeq_{\pK\rightarrow \xi}=\left(\xi,1,\text{--},\text{--},\text{--},\text{--}\right)$.}
\begin{align}
\pSeq_\rv{t}
%&=
%\eOP{1}{\tfrac{1}{\omega_1\omega_2}}{-\left(1+\omega_1\omega_2\right)}{\pSeq_{\left(\omega_1\right)}^{-1}}
%\nonumber \\
&=
	\Bigl(
		\tfrac{\pK_{\left(\omega_1\right)} }{{\pC_{\left(\omega_1\right)} }},
		1,
		%\left(1-\bV_{{\left(\omega_1\right)}}-\left(1+\frac{1}{\omega_1\omega_2}\right)\tV_{{\left(\omega_1\right)}}\right),
		\left(\B{1}_{\pQ_1+\pQ_2}-\bV_{\left(\omega_1\right)}-\tV_{\left(\omega_1\right)}-\tfrac{1}{\omega_1\omega_2}\tV_{\left(\omega_1\right)}\right),
			\nonumber \\
	&\hspace{3cm}
		\left(\B{1}_{\pP_1+\pP_2}-\aV_{\left(\omega_1\right)}-\sV_{\left(\omega_1\right)}-\tfrac{1}{\omega_1\omega_2}\sV_{\left(\omega_1\right)}\right), 
		\tfrac{1}{\omega_1\omega_2}\tV_{\left(\omega_1\right)},
		\tfrac{1}{\omega_1\omega_2}\sV_{\left(\omega_1\right)}
	\Bigr).
\end{align} 
%

%and its density function $\PDF{\rv{t}}{t}$ is given by
%
%\begin{align}	\label{eq:pdf:fpt}
%\PDF{\rv{t}}{t}
%&=
%	\frac{2}{a^{1/\left(\omega_1\omega_2\right)}}
%	\Fox{\pM_1+\pM_2}{\pN_1+\pN_2}{\pP_1+\pP_2}{\pQ_1+\pQ_2}
%	{
%		\frac{
%				a^{1/\left(\omega_1\omega_2\right)}}{
%			t};
%		\eOP{1}{\frac{1}{\omega_1\omega_2}}{1+\omega_1\omega_2}{\pSeq_{\left(\omega_1\right)}}
%	}.
%\end{align}
\begin{proof}
With the absorbing boundary condition $\PDF{\rp{x}{t}}{a}=0$, the density function of the position of molecule $x$ at time $t$, denoted by $\PDF{\rp{\tilde{x}}{t}}{x}$ for $x<a$, can be found using the image method \cite[Section~3.2]{Red:01:Book} as
\begin{align}	\label{eq:barw}
	\PDF{\rp{\tilde{x}}{t}}{x}
	&=
		\PDF{\rp{x}{t}}{x}-\PDF{\rp{x}{t}}{-x+2a}.
\end{align}
Let 
$
S_{\rv{t}}\left(t\right)\triangleq
		\int_{-\infty}^{a}
		\PDF{\rp{\tilde{x}}{t}}{x}
		dx
$
be the \emph{survival probability} that the molecule is located at $x<a$ for all times up to $t$. Then, we have
\begin{align}
	\PDF{\rv{t}}{t}
	&\overset{\text{(a)}}{=}
		-\frac{\partial S_{\rv{t}}\left(t\right)}{\partial t}
	\nonumber \\
	&
	\overset{\text{(b)}}{=}
		-
		\frac{\partial}{\partial t}
		\int_{-\infty}^{a/t^{\omega_1\omega_2}}
		\PDF{\sNorm{\rp{x}{1}}}{y}
		dy
		+
		\frac{\partial}{\partial t}
		\int_{a/t^{\omega_1\omega_2}}^{\infty}
		\PDF{\sNorm{\rp{x}{1}}}{y}
		dy
	\nonumber \\
	&\overset{\text{(c)}}{=}
		-2
		\frac{\partial}{\partial t}
		\int_{0}^{a/t^{\omega_1\omega_2}}
		\PDF{\sNorm{\rp{x}{1}}}{y}
		dy
	\nonumber \\
	&
	=
		\frac{2a\omega_1\omega_2}{t^{\omega_1\omega_2+1}}
		\PDF{\sNorm{\rp{x}{1}}}{\frac{a}{t^{\omega_1\omega_2}}}
\end{align}
where (a) follows from the relationship between the FPT and survival probability as $\PDF{\rv{t}}{t}dt=S_\rv{t}\left(t\right)-S_\rv{t}\left(t+dt\right)$ with the first-order Taylor approximation \cite{Red:01:Book}; (b) is obtained from the self-similar property of $\rp{x}{t}$ in \eqref{eq:barw} and by making changing the  variable $x/t^{\omega_1\omega_2}=y$; and (c) follows from the fact that $\PDF{\sNorm{\rp{x}{1}}}{x}$ has a symmetric relation. Finally, using the modulation property of $H$-function \cite[Table~II]{JSW:15:IT}, the desired result \eqref{eq:foxv:fpt} is obtained. %and \eqref{eq:pdf:fpt} are obtained.
\end{proof}
\end{theorem}

\begin{definition}[$H$-Noise]	\label{def:Hnoise}
The FPT $\rv{t}$ acts as the \emph{unique} source of uncertainty or distortion in molecular communication. We refer to the FPT of $H$-diffusion molecules in Theorem~\ref{thm:fpt:Hdiffusion} as the \emph{$H$-noise}. Specifically, we define the FPT of SHD molecules in Definition~\ref{def:StdHdiffuion} as the \emph{standard $H$-noise}. 
\end{definition}

\begin{remark}[Standard $H$-noise] 
The  standard $H$-noise, denoted by $\rv{t}_\mathrm{sHn}$, is the $H$-variate
\begin{align}
\rv{t}_\mathrm{sHn}
\sim
	\FoxV{1}{2}{3}{3}
	{\pSeq_\mathrm{sHn}\Ket{\left(\frac{a}{\beta_1\beta_2^{\omega_1}}\right)^{1/\left(\omega_1\omega_2\right)}}}
\end{align}
where the parameter sequence $\pSeq_\mathrm{sHn}$ is
\begin{align}
\pSeq_\mathrm{sHn}
&=
\Bigl(
	\tfrac{2}{\alpha_1},
	1,
	\left(-\tfrac{1}{\omega_2},-\tfrac{1}{\omega_1\omega_2},-\tfrac{1}{2\omega_1\omega_2}\right),
	\left(-\tfrac{1}{\alpha_1\omega_1\omega_2},-\tfrac{1}{2\omega_1\omega_2},-\tfrac{\alpha_2}{\omega_2}\right),
	\nonumber \\
&\hspace{5.5cm}
	\left(\tfrac{1}{\omega_2},\tfrac{1}{\omega_1\omega_2},\tfrac{1}{2\omega_1\omega_2}\right),
	\left(\tfrac{1}{\alpha_1\omega_1\omega_2},\tfrac{1}{2\omega_1\omega_2},\tfrac{\alpha_2}{\omega_2}\right)
\Bigr).
\end{align}
\end{remark}

\subsection{$H$-Noise Power}

Since the \emph{algebraic-tailed} random variable $\rv{z}$ exhibits finite absolute moments only for the order $\ell$ less than $\eta$, i.e., $\mathbb{E}\bigl\{\left|\rv{z}\right|^\ell\bigr\}< \infty$ for $\ell<\eta$, the second moment that is widely used as a standard signal power measure does not exist for the class of algebraic distributions when $\eta<2$. The fractional lower-order statistics (FLOS) theory is a useful tool to measure and characterize the behavior of impulsive signals with an algebraic distribution \cite{MN:96:SP}, but it cannot provide a universal framework for characterization of algebraic-tailed processes. 
In this subsection, we provide a concept of \emph{$H$-noise power} using the finite logarithm moments based on zero-order statistics \cite[Theorem~1]{GPA:06:SP}.

\begin{theorem}[$H$-Noise Tails]	\label{thm:Hnoise:tails}
Let 
\begin{align}	\label{def:Hnoise:tails}
	\kappa
	\triangleq
	\lim_{t\rightarrow \infty}
	\frac{
		-\log S_{\rv{t}}\left(t\right)
		}
		{
		\log t}
\end{align}
be the tail constant of $\rv{t}$. Then, we have
\begin{align}
\kappa
=
	\omega_1\omega_2\cdot
	\min_{j=1,\ldots,\pM_1+\pM_2}
	\left\{
			1+\frac{\Re\left(\bV_{\left(\omega_1\right),j}\right)}{
			\tV_{{\left(\omega_1\right)},j}
			}
	\right\}
\end{align}
where $\bV_{\left(\omega_1\right),j}$ and $\tV_{{\left(\omega_1\right)},j}$ are the $j$th element of the sequences $\bV_{\left(\omega_1\right)}$ and $\tV_{{\left(\omega_1\right)}}$, respectively.
\begin{proof}
Using the formulation for the CDF of $H$-variate \cite[Table~VI]{JSW:15:IT}, we have
\begin{align}	\label{eq:cdf:Hn}
	S_{\rv{t}}\left(t\right)
	&=
	\Prob{\rv{t}>t}
	\nonumber \\
	%&
&=
		1-\FoxHT{\pN_1+\pN_2,\pM_1,\pM_2}{\pQ_1+\pQ_2,\pP_1+\pP_2}
		{
			\Bra{1}
			\left(\pSeq_\rv{t}\Ket{\left(a\pC_{\left(\omega_1\right)}\right)^{1/\left(\omega_1\omega_2\right)}}\right)
		}
		{\frac{1}{t}\IndF{\left[0,1\right]}{t}}
		{t}
%	\nonumber \\
%	&=
%		\Fox{\pN_1+\pN_2+1}{\pM_1+\pM_2}{\pQ_1+\pQ_2+1}{\pP_1+\pP_2+1}
%		{t;\canOP{\pSeqcdf^{-1}}
%			{
%				\Bra{1}
%			\left(\pSeq_\rv{t}\Ket{\left(a\pC_{\left(\omega_1\right)}\right)^{1/\left(\omega_1\omega_2\right)}}\right)
%			}
%		}
	\nonumber \\
	&=
		\Fox{\pN_1+\pN_2+1}{\pM_1+\pM_2}{\pQ_1+\pQ_2+1}{\pP_1+\pP_2+1}
		{\frac{t}{\left(a\pC_{\left(\omega_1\right)}\right)^{1/\left(\omega_1\omega_2\right)}};\canOP{\pSeqcdf^{-1}}
			{
				\Bra{1}
				\pSeq_\rv{t}
			}
		}.
\end{align}
Then, using the algebraic asymptotic expansion of the $H$-function \cite[Proposition~3]{JSW:15:IT}, we get
\begin{align}
S_{\rv{t}}\left(t\right)
&\doteq
		\Fox{\pN_1+\pN_2+1}{\pM_1+\pM_2}{\pQ_1+\pQ_2+1}{\pP_1+\pP_2+1}
		{t;\canOP{\pSeqcdf^{-1}}
			{
				\Bra{1}
				\pSeq_\rv{t}			
			}}
\nonumber\\
&
\doteq
		t^{-\omega_1\omega_2\cdot
		\min_{j=1,\ldots,\pM_1+\pM_2}
		\left\{
			1+
			\frac{\Re\left(\bV_{{\left(\omega_1\right)},j}\right)}{
			\tV_{{\left(\omega_1\right)},j}
			}
		\right\}
		}
\end{align}
which completes the proof.
\end{proof}
\end{theorem}

\begin{remark}[Algebraic-Tailed or Heavy-Tailed Distribution]
Since $S_{\rv{t}}\left(t\right)$ (also called a tail function $\Prob{\rv{t}>t}$) in Theorem~\ref{thm:Hnoise:tails} has a polynomial decay rate $\kappa \in \R_{++}$, the $H$-noise can be said to be an algebraic-tailed random variable \cite{Arc:04:Book}. In addition, since all algebraic-tailed random variables possess heavier tails than a family of exponential distributions, we can also call it the heavy-tailed distribution. Hence, the $H$-noise $\rv{t}$ has a finite moment $\EX{\rv{t}^\ell}$ for $\ell < \kappa$.
\end{remark}

\begin{remark}[Standard $H$-Noise Tails] 
The tail constant for the standard $H$-noise $\kappa_\mathrm{sHn}$ is 
\begin{align}
\kappa_\mathrm{sHn}
&=
	\begin{cases}
	\omega_1\omega_2, & \omega_1 < 1 \\
	\omega_2, & \omega_1 \geq 1.
	\end{cases}
\end{align}
\end{remark}

\begin{theorem}[Logarithm Moments of $H$-Noise]	\label{thm:LM:Hnoise}
Note that any algebraic-tailed distribution $\rv{x}$ has a finite logarithm moment \cite[Theorem~2.5]{Arc:04:Book}, i.e., $\EX{\ln\left(\rv{x}\right)}<\infty$. Hence, the logarithm moment of $H$-noise $\rv{t}$ exists for all ranges of parameters and is given by
\begin{align}	\label{eq:LM:Hnoise:HT}
	\EX{\ln\left(\rv{t}\right)}
	&=
		\FoxHT{2,2}{2,2}
		{\pSeq_\mathrm{ln}}
		{\left(t-1\right)\PDF{\rv{t}}{t}}
		{1}
	\\ \label{eq:LM:Hnoise:HF}
	&=
		\Fox{\pM_1+\pM_2+2}{\pN_1+\pN_2+2}{\pP_1+\pP_2+2}{\pQ_1+\pQ_2+2}
		{1;\stdOP{\pSeq_\mathrm{ln}}{\Bra{1\frac{}{}}\left(\pSeq_\rv{t}}\Ket{\left(a\pC_{\left(\omega_1\right)}\right)^{\frac{1}{\omega_1\omega_2}}}\right)}
	\nonumber \\
	&\hspace{2.5cm}
		-
		\Fox{\pM_1+\pM_2+2}{\pN_1+\pN_2+2}{\pP_1+\pP_2+2}{\pQ_1+\pQ_2+2}
		{\left(a\pC_{\left(\omega_1\right)}\right)^{\frac{1}{\omega_1\omega_2}};\stdOP{\pSeq_\mathrm{ln}}{\pSeq_\rv{t}}}
\end{align} 
where $\pSeq_\mathrm{\ln}=\left(1,1,\left(\B{0}_2,\text{--}\right),\left(\B{0}_2,\text{--}\right),\left(\B{1}_2,\text{--}\right),\left(\B{1}_2,\text{--}\right)\right)$. For standard $H$-noise, the logarithm moment of $H$-noise in \eqref{eq:LM:Hnoise:HF} reduces to
\begin{align}	\label{eq:LM:stdHnoise}
	\EX{\ln\left(\rv{t}_\mathrm{sHn}\right)}
	=
		\left(
			\frac{
			1-1/\alpha_1+\left(1-\alpha_2\right)\omega_1}{
			\omega_1\omega_2}
		\right)
		\gamma_\mathrm{e}
		+
		\frac{1}{\omega_1\omega_2}
		\ln\left(\frac{a}{\beta_1\beta_2^{\omega_1}}\right)
\end{align}
where $\gamma_\mathrm{e}\approx 0.57721$ is the Euler-Mascheroni constant.
\begin{proof}
Using \eqref{eq:foxv:fpt} and \cite[Example~4]{JSW:15:IT}, the $H$-transform expression is obtained for the logarithm moment of $H$-noise \eqref{eq:LM:Hnoise:HT}, from which along with the Mellin operation \cite[Proposition~4]{JSW:15:IT}, we arrive at the desired result \eqref{eq:LM:Hnoise:HF}. For standard $H$-noise, using the relation between logarithm moment and derivative of moment such that
\begin{align}
	\EX{\ln\left(\rv{t}_\mathrm{sHn}\right)}
	=
		\left.\frac{\partial \EX{\rv{t}_\mathrm{sHn}^\ell}}{\partial\ell}\right|_{\ell=0}
\end{align}
we obtained \eqref{eq:LM:stdHnoise}, which completes the proof.
\end{proof}
\end{theorem}

\begin{corollary}[Geometric Power of Standard $H$-Noise]	\label{cor:gmp:Hnoise}
Let 
\begin{align}
	\GMP{\rv{t}}
	\triangleq
	\exp\left\{\EX{\ln\left(\rv{t}\right)}\right\}
\end{align}
be the \emph{geometric power} of random variable $\rv{t}$. Then, the geometric power of $H$-noise $\GMP{\rv{t}}$ can be obtained generally using the logarithm moments of $H$-noise given in \eqref{eq:LM:Hnoise:HT}. Specifically, for the standard $H$-noise, $\GMP{\rv{t}_\mathrm{sHn}}$ has a compact form of
\begin{align}
	\GMP{\rv{t}_\mathrm{sHn}}
	=
		\left(\frac{a\mathcal{G}^{1-1/\alpha_1+\left(1-\alpha_2\right)\omega_1}}{\beta_1\beta_2^{\omega_1}}\right)^{\frac{1}{\omega_1\omega_2}}
\end{align}
where $\mathcal{G}\triangleq e^{\gamma_{\mathrm{e}}}\approx 1.78107$ denotes the exponential Euler-Mascheroni constant.\footnote{The measure of geometric power was introduced in \cite{GPA:06:SP} for the processing and characterization of very impulsive signals with the concept of zero-order statistics. Specifically, the symmetric $\alpha$-stable distribution $\rv{s} \sim \Stable{\Salpha}{0}{\Sgamma}{0}$ was considered where its geometric power was shown as $\GMP{\rv{s}} = \left(\mathcal{G}\Sgamma\right)^{1/\Salpha}/\mathcal{G}$.} 
\begin{proof}
It follows readily from Theorem~\ref{thm:LM:Hnoise}.
\end{proof}

\end{corollary}

\begin{remark}[Geometric Mean, Power, and FLOS]	\label{rem:GPF}
The geometric power $\GMP{\rv{t}}$ has a relation to the geometric mean of nonnegative random variable $\rv{t}$ by %the fact that
\begin{align}
\GMP{\rv{t}}
&=\exp\left\{ \lim_{N \rightarrow \infty}\frac{1}{N}\sum_{i=1}^{N}\ln\left(t_i\right)
\right\}
\nonumber \\
&
=\lim_{N \rightarrow \infty}\left\{\prod_{i=1}^{N} t_i\right\}^{1/N}
\end{align}
where $\left(t_1, \ldots, t_N\right)$ is a sequence of independent samples initiated by random variable $\rv{t}$. Compared to the arithmetic mean, the geometric mean is said to be not overly influenced by the very large values in a skewed distribution. This advantage is appropriate for $H$-noise. Since the geometric power is linked to the geometric mean, we use the square of $\GMP{\rv{t}}$ for the \emph{$H$-noise power}, denoted by $\mathcal{N}\left(\rv{t}\right)=\left\{\GMP{\rv{t}}\right\}^2$.\footnote{
The geometric power also can be linked to the FLOS method if there exists a sufficiently small value $\ell$ satisfying \cite{GPA:06:SP}
$
\GMP{\rv{t}}
	=\lim_{\ell \rightarrow 0} \left\{\EX{\rv{t}^\ell}\right\}^{1/\ell}
$. 
This reveals that the geometric power can be used mathematically and conceptually in a rich set of heavy-tailed distributions.
}
\end{remark}

\begin{remark}[Normal Diffusion]
The $H$-noise $\rv{t}$ in Brownian motion without drift has a nonnegative stable distribution with the characteristic exponent $1/2$ (L{\'e}vy distribution) where the PDF $\PDF{\rv{t}}{t}$ is given by\footnote{The $H$-noise in Brownian motion with nonzero drift follows an inverse Gaussian distribution \cite{SEA:12:IT}.}
\begin{align}
	\PDF{\rv{t}}{t}
	&=
	\frac{4}{a^2\sqrt{\pi}}\FoxH{0}{1}{1}{0}{\frac{4t}{a^2}}{\left(-\frac{1}{2},1\right)}{\text{---}}
	\nonumber \\
	&
	=
	\frac{a}{\sqrt{4\pi t^3}}\exp\left(-\frac{a^2}{4t}\right)
\end{align}
and its corresponding geometric power $\GMP{\rv{t}}$ is given by $\GMP{\rv{t}} = a^2\mathcal{G}$ \cite{FMGCEG:19:MBSC}. 
%\begin{align}	\label{eq:gmp:st}
%\GMP{\rv{t}} &= a^2\mathcal{G}.
%\end{align} 
%
Table~\ref{table:Hnoise} shows the $H$-noise $\rv{t}$ and its geometric power $\GMP{\rv{t}}$ for the typical anomalous diffusion models in Table~\ref{table:TDM:HD}.
\end{remark}

%%%%%%%%%%%%%%%%%%%%%%%%%%%%%%%%%%%%%%%%
\begin{table}[t]
\caption{$H$-Noise $\rv{t}$ and Its Geometric Power $\GMP{\rv{t}}$ for Typical Anomalous Diffusion 
in Table~\ref{table:TDM:HD}:
$$
	\rv{t}\sim\FoxD{\oSeq,\pSeq\Ket{a^{1/\omega}}},\quad 
	\GMP{\rv{t}}=a^{1/\omega}\mathcal{G}^{1/\omega-c}
$$
} \centering
\label{table:Hnoise}

\begin{threeparttable}
\begin{adjustbox}{max width=0.9\textwidth}
\begin{tabular}{lllll}
%\toprule
\midrule
\midrule
Diffusion &
\multicolumn{2}{l}{$H$-noise $\rv{t} \sim \FoxD{\oSeq,\pSeq}$} & %\sim\FoxD{\oSeq,\pSeq\Ket{a^{\omega}}}$} &
\multicolumn{2}{l}{Geometric Power $\GMP{\rv{t}}$}
\\[-0.1cm]

$\rp{h}{t}$ &
$\oSeq$ &%=\left(\pM_1,\pN_1,\pP_1,\pQ_1\right)$ &
$\pSeq$ &%=\left(\pK_1,\pC_1,\aV_1,\bV_1,\sV_1,\tV_1\right)$ &
$\omega$ &
$c$
\\

%\midrule
\midrule

ST-FD &
$\left(1,2,3,3\right)$ &
$\left(
	\frac{2}{\alpha}, 1,\left(-\frac{1}{\beta},-\frac{\alpha}{\beta},-\frac{\alpha}{2\beta}\right),\left(-\frac{1}{\beta},-\frac{\alpha}{2\beta},-1\right),\left(\frac{1}{\beta},\frac{\alpha}{\beta},\frac{\alpha}{2\beta}\right),\left(\frac{1}{\beta},\frac{\alpha}{2\beta},1\right)
\right)$ &
$\beta/\alpha$ &
$1$
\\ \midrule

S-FD &
$\left(1,1,2,2\right)$ &
$\Bigl(
	\frac{2}{\alpha}, 1,\left(-\alpha,-\frac{\alpha}{2}\right),\left(-1,-\frac{\alpha}{2}\right),\left(\alpha,\frac{\alpha}{2}\right),\left(1,\frac{\alpha}{2}\right)	
\Bigr)$ &
$1/\alpha$ &
$1$
\\ \midrule

T-FD &
$\left(0,1,1,1\right)$ &
$\Bigl(
	1, 1,-\frac{2}{\beta},-1,\frac{2}{\beta},1
\Bigr)$ &
$\beta/2$ &
$1$
\\ \midrule

EK-FD &
$\left(0,2,2,1\right)$ &
$\Bigl(\frac{4^{1/\alpha}}{\sqrt{\pi}},4^{1/\alpha},\left(-\frac{1}{\alpha},\frac{1}{2}-\frac{1}{\alpha}\right),-\frac{\beta}{\alpha},\frac{1}{\alpha}\B{1}_2,\frac{\beta}{\alpha}
\Bigr)$ &
$\alpha/2$ &
$\beta/\alpha$
\\ \midrule

GBM &
$\left(0,2,2,1\right)$ &
$\Bigl(\frac{4^{1/\beta}}{\sqrt{\pi}},4^{1/\beta},\left(-\frac{1}{\beta},\frac{1}{2}-\frac{1}{\beta}\right),-1,\frac{1}{\beta}\B{1}_2,1
\Bigr)$ &
$\beta/2$ &
$1$
\\ \midrule

FBM &
$\left(0,1,1,0\right)$ &
$\Bigl(\frac{4^{1/\alpha}}{\sqrt{\pi}},4^{1/\alpha},\frac{1}{2}-\frac{1}{\alpha},\text{--},\frac{1}{\alpha},\text{--}
\Bigr)$ &
$\alpha/2$ &
$1/\alpha$
\\ \midrule

BM &
$\left(0,1,1,0\right)$ &
$\Bigl(\frac{4}{\sqrt{\pi}},4,-\frac{1}{2},\text{--},1,\text{--}\Bigr)$ &
$1/2$ &
$1$
\\
\midrule
\midrule
%\bottomrule
\end{tabular}
\end{adjustbox}
\end{threeparttable}

\end{table}

%\fi %%%%%%%%%%%%%%%%%%%%%%%%%%%%%%%%%%%%%%%%%%

\subsection{Numerical Examples}

In what follows, we use $\left(\alpha_1,\alpha_2\right)$-SHD to denote the particularized SHD where the diffusion parameters are $\omega_1=1/\alpha_1$, $\omega_2=\alpha_2$, and $\beta_1\beta_2^{1/\alpha_1}=K^{1/{\alpha_1}}$. To exemplify the different types of diffusion scenarios, we consider:
       i) $\left(\alpha_1,\alpha_2\right)=\left(2,1\right)$ for normal diffusion; 
       ii) $\left(\alpha_1,\alpha_2\right)=\left(2,0.5\right)$ for subdiffusion; and  
       iii) $\left(\alpha_1,\alpha_2\right)=\left(1.8,1\right)$ for superdiffusion.
%\footnote{The diffusion coefficient $10^{-10}$~[$\mathrm{m^2/s}$] is of a biological environment \cite{PA:14:COM}. The various examples for diffusion coefficient can be founded in \cite{BS:06:Book}, for example, $10^{-9}$~[$\mathrm{m^2/s}$] for self diffusion and $10^{-11}$~[$\mathrm{m^2/s}$] for colloidal substances.}  
 
 \begin{figure}[t!]
    \centerline{\includegraphics[width=0.55\textwidth]{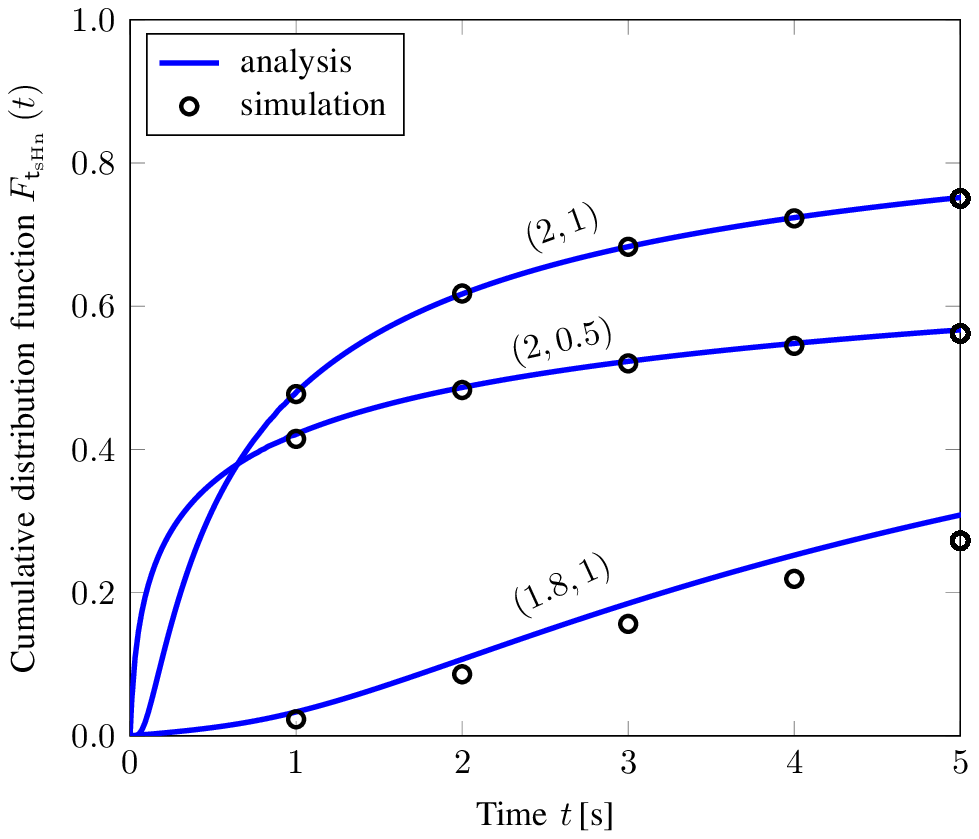}}
    %\vspace{-0.3cm}
\caption{
       CDF $\CDF{\rv{t}_{\mathrm{sHn}}}{t}$ of the standard $H$-noise $\rv{t}_\mathrm{sHn}$ in the $\left(\alpha_1,\alpha_2\right)$-SHD at the distance $a=10^{-5}$~[m] for:  
       i) $\left(\alpha_1,\alpha_2\right)=\left(2,1\right)$; 
       ii) $\left(\alpha_1,\alpha_2\right)=\left(2,0.5\right)$; and  
       iii) $\left(\alpha_1,\alpha_2\right)=\left(1.8,1\right)$;
       with $K=10^{-10}$~[m$^2$/s].
}
\label{fig:6}
\end{figure}
\begin{figure}[t!]
    \centerline{\includegraphics[width=0.55\textwidth]{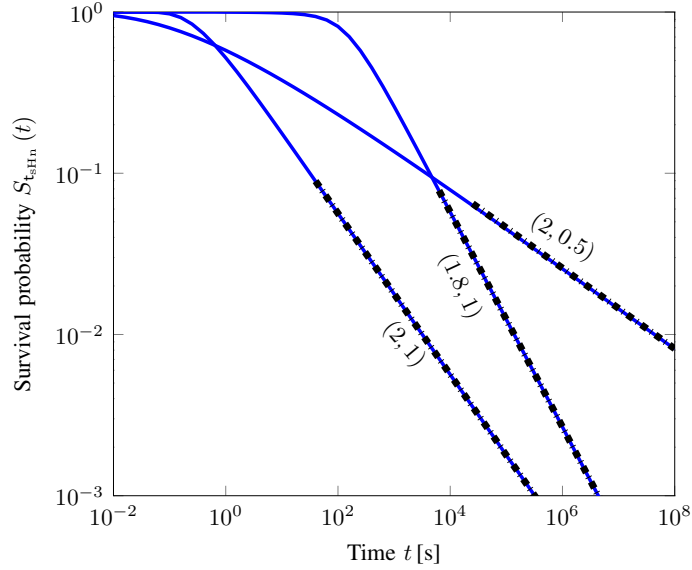}}
    %\vspace{-0.3cm}
\caption{
       Survival probability $S_{\rv{t}_{\mathrm{sHn}}}\left(t\right)$ of the standard $H$-noise $\rv{t}_\mathrm{sHn}$ in the $\left(\alpha_1,\alpha_2\right)$-SHD for: 
       i) $\left(\alpha_1,\alpha_2\right)=\left(2,1\right)$; 
       ii) $\left(\alpha_1,\alpha_2\right)=\left(2,0.5\right)$; and  
       iii) $\left(\alpha_1,\alpha_2\right)=\left(1.8,1\right)$;
       with $K=10^{-10}$~[m$^2$/s]. The black dashed-dotted line stands for the tail constant $\kappa$ of $H$-noise in~\eqref{def:Hnoise:tails}.
       }
\label{fig:7}
\end{figure}

Fig.~\ref{fig:6} shows the CDF $\CDF{\rv{t}_{\mathrm{sHn}}}{t}$ of the standard $H$-noise $\rv{t}_\mathrm{sHn}$ in the $\left(\alpha_1,\alpha_2\right)$-SHD at distance $a=10^{-5}$~[m]. We can observe that the anomalous diffusions for $\left(\alpha_1,\alpha_2\right)=\left(2,0.5\right)$ and $\left(\alpha_1,\alpha_2\right)=\left(1.8,1\right)$ have a large dispersion in propagation compared to the normal diffusion for $\left(\alpha_1,\alpha_2\right)=\left(2,1\right)$.\footnote{The simulation is conducted based on the particle-based simulator \cite{CTJS:15:CL, TJS:19:ACCESS, TJSW:18:COM, AJS:18:COM}. The simulation of FPT in superdiffusion is overestimated due to the long jump property of  molecules \cite{TJSW:18:COM}. } To demonstrate the heavy-tailed property of $H$-noise, the survival probability $S_{\rv{t}_{\mathrm{sHn}}}\left(t\right)$ of the standard $H$-noise $\rv{t}_\mathrm{sHn}$ in $\left(\alpha_1,\alpha_2\right)$-SHD for the three diffusion scenarios  is depicted in Fig.~\ref{fig:7}. It can be seen from the figure that the $H$-noise distribution follows the asymptotic tail constant scaling behaviours as discussed in Theorem~\ref{thm:Hnoise:tails}. In this example, the tail constants (slope of black dashed-dotted line in the figure) are equal to $\kappa=0.5$, $0.25$, and $0.56$ for $\left(2,1\right)$-, $\left(2,0.5\right)$-, and $\left(1.8,1\right)$-SHD, respectively.

\begin{figure}[t!]
    \subfigure[$a=10^{-5}~\text{[m]}$]{
        \centerline{\includegraphics[width=0.6\textwidth]{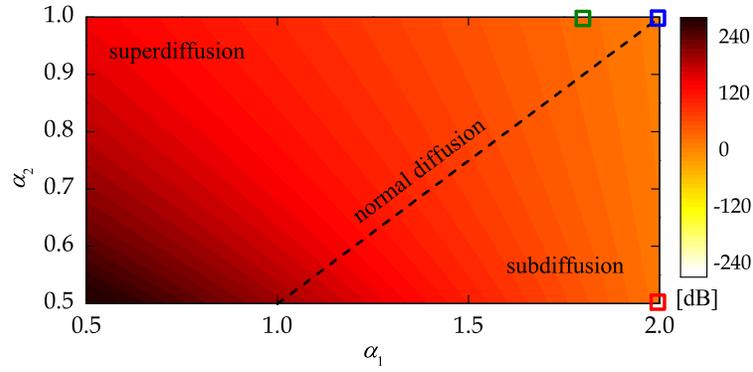}}
        \label{fig:8:a}
    }\hfill
    \subfigure[$a=10^{-8}~\text{[m]}$]{
        \centerline{\includegraphics[width=0.6\textwidth]{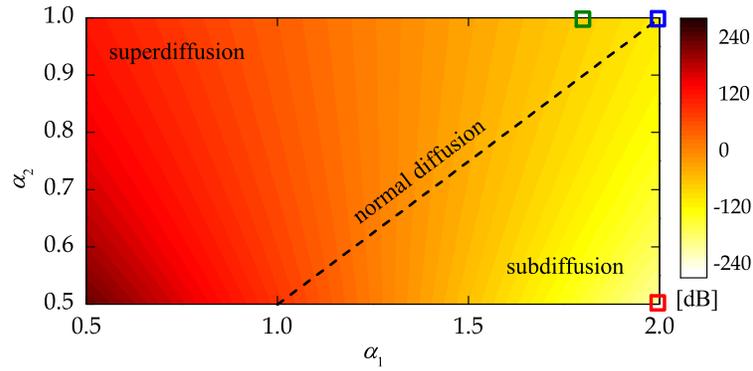}}
        \label{fig:8:b}
    }\hfill
    \subfigure[$a=10^{-10}~\text{[m]}$]{
        \centerline{\includegraphics[width=0.6\textwidth]{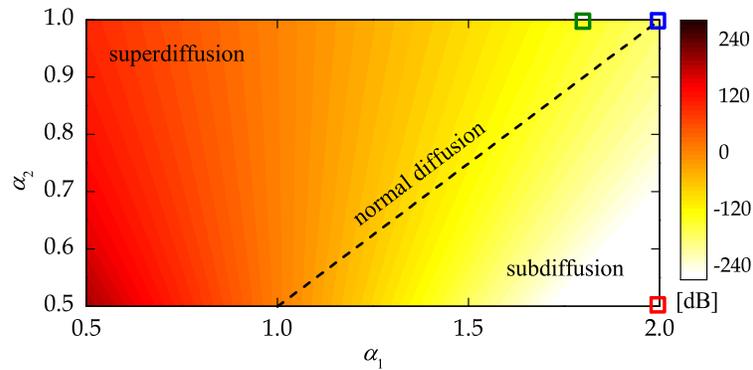}}
        \label{fig:8:c}
    }\hfill    
    \caption{
       $H$-noise power $\mathcal{P}\left(\rv{t}_\mathrm{sHn}\right)$ [dB] in the $\left(\alpha_1,\alpha_2\right)$-SHD as a function of $\left(\alpha_1,\alpha_2\right)$ at the distance (a) $a=10^{-5}$~[m], (b) $a=10^{-8}$~[m], and (c) $a=10^{-10}$~[m] with $K=10^{-10}$~[m$^2$/s].
    }
    \label{fig:8}
\end{figure}

Fig.~\ref{fig:8} shows the $H$-noise power  $\mathcal{N}\left(\rv{t}_\mathrm{sHn}\right)$ [dB] in the $\left(\alpha_1,\alpha_2\right)$-SHD as a function of $\left(\alpha_1,\alpha_2\right)$ at distance (a) $a=10^{-5}$~[m], (b) $a=10^{-8}$~[m], and (c) $a=10^{-10}$~[m] with $K=10^{-10}$~[m$^2$/s]. The three diffusion scenarios are indicated as the blue square at $\left(\alpha_1,\alpha_2\right)=\left(2,1\right)$ for normal diffusion, the red square at $\left(\alpha_1,\alpha_2\right)=\left(2,0.5\right)$ for subdiffusion, and the green square at $\left(\alpha_1,\alpha_2\right)=\left(1.8,1\right)$ for superdiffusion. Given a fixed diffusion coefficient, the $H$-noise power increases with large distance $a$ and low value of $\alpha_1$. However, with fixed $\alpha_1$, the noise power decreases with $\alpha_2$ when $a^{\alpha_1}/K > 1$ (Fig.~\ref{fig:8:a}), while it increases in the opposite case (Fig.~\ref{fig:8:b} and Fig.~\ref{fig:8:c}). Thus, the error performance in $\left(2,0.5\right)$-SHD outperforms that of $\left(2,1\right)$-SHD in the low-SNR regime (see also Fig.~\ref{fig:10}).  Similarity, with fixed $\alpha_2$, the noise power decreases with $\alpha_1$ when $a<1$ (Fig.~\ref{fig:8}), while it increases with $\alpha_1$ when $a>1$. This leads to intersection of the BER curves in $\left(2,1\right)$-SHD and $\left(1.8,1\right)$-SHD as in Fig.~\ref{fig:10}.

\section{Error Probability Analysis}

In this section, we characterize the effect of $H$-diffusion on the error performance of molecular communication. Specifically, we consider an $M$-ary transmission scheme to boost the data rate as well as a  $N$-molecule transmission scheme to increase reliability for molecular communication.

\subsection{Molecular Communication System Model}

\begin{figure}[t]
    \centerline{\includegraphics[width=0.95\textwidth]{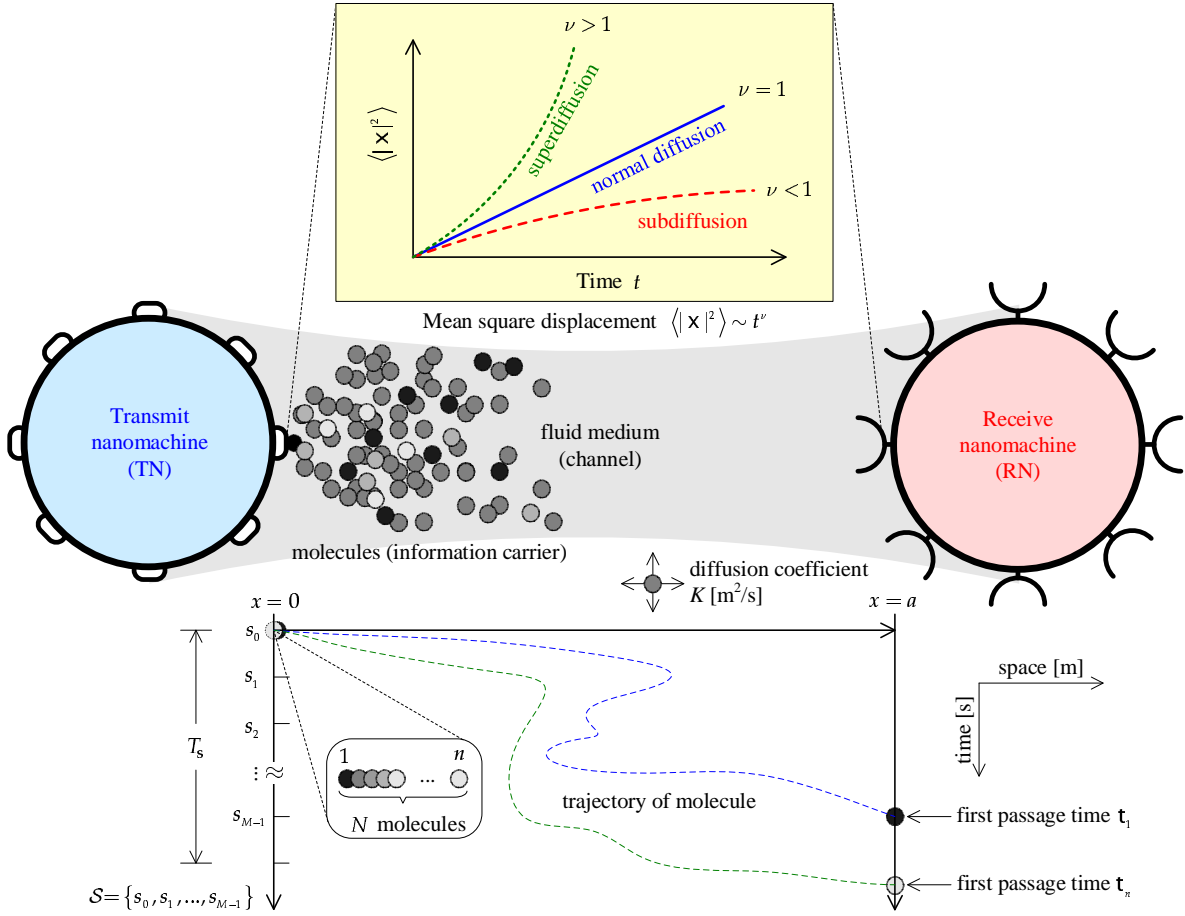}}
    \caption{
        Molecular communication system: a TN emits an information molecule to a RN in a fluid medium where the motion of information molecule is determined by the relationship between mean square displacement and time.
    }
    \label{fig:9}
\end{figure}

We consider a molecular communication system, as illustrated in Fig.~\ref{fig:9}, where a TN located at $x=0$ emits molecules (information carrier) to an RN located  $a$~[m] from the TN. The emitted information molecules are assumed to be randomly and freely propagated in the fluid medium, e.g., blood vessels or tissues, under the $H$-diffusion laws with the diffusion coefficient $K$~[m$^2$/s]. In this paper, we consider an molecular communication system with the following assumptions: 1) the TN can perfectly control the releasing time and the number of molecules for each symbol message; 2) the clock of the TN is perfectly synchronized with that at the RN; 3) movements of each molecule in the fluid medium are independent and identically distributed; 4) the RN acts as a perfectly absorbing boundary and perfectly measures the arrival time of molecules; 5) the molecules that arrived at the RN are absorbed and removed from the system; 6) the TN uses different types of molecule for each symbol to avoid inter-symbol interference.\footnote{The number of molecule types used for this model can be minimized by introducing the lifetime of molecules \cite{SML:15:WCOM,TJSW:18:COM} and the chemical reactions in the medium \cite{AMGMKF:17:MBSC,MGMK:18:COM,FMGCEG:19:MBSC}.}
%6) the RN can wait for a infinite waiting time; and 7) inter symbol time is large enough to avoid inter symbol interference. 

The information is encoded based on the release time. Let $\mathcal{S}\triangleq \left\{s_0, s_1, \ldots, s_{M-1} \right\}$ be the set of molecular release times  at the TN for corresponding symbol constellation where $s_i = i T_\mathrm{s}/M$, $M$ is a modulation order, and $T_\mathrm{s}$ is a symbol time. The TN is able to emit $N$ molecules for a one-symbol transmission at the release time $\rv{s} \in \mathcal{S}$. Hence, for the $n$th molecule among $N$ emitted molecules at the release time $\rv{s}$, the arrival time $\rv{y}_n$ at the RN is
\begin{align}	\label{eq:at}
\rv{y}_n=\rv{s}+\rv{t}_{n}
\end{align}
where $\rv{t}_{n}$ is the $H$-noise of the $n$th molecule. Then, the transmitted symbol $\rv{s}$ can be decoded using the set of arrival times $\mathcal{Y}=\left\{\rv{y}_1, \rv{y}_2, \ldots,\rv{y}_N\right\}$ at the RN. Since $H$-noise can be the heavy-tailed random variable, it may have a large waiting time for molecules arriving at the RN, i.e., there exists a positive probability that the molecule will not have arrived at TN within finite time. Furthermore, with the large number of released molecules, the RN needs to wait until all molecules are absorbed. Hence, we consider a \emph{first arrival detection} that uses the time of first arrival molecule at the RN among $N$ released molecules to decode the transmitted symbol.\footnote{Several detection schemes are proposed in \cite{MFCG:16:GLOBECOM} to decode the transmitted signal using multiple transmitted molecules. In this work, we use the simple detection scheme because the complex implementation for a signal detection using multiple arrival molecules is impractical at the biological circuit level \cite{AFSFH:12:MWCOM,NSOMV:14:NB}.}. Then, the explicit signal model for a one-symbol transmission is 
\begin{align}	\label{eq:at2}
\rv{y}=\rv{s}+\rv{t}_\mathrm{min}
\end{align}
where 
$
\rv{t}_\mathrm{min}=\min\left\{\rv{t}_{1},\rv{t}_{2},\ldots,\rv{t}_{N}\right\}
$ 
is referred as the \emph{first arrival $H$-noise}.

\subsection{Error Probability Analysis}

The information can be decoded using the maximum likelihood detection for $M$-ary modulation:
\begin{align}
%\displaybreak
\hat{\rv{s}}
&=
	\argmax_{\rv{s}=\left\{s_0,s_1,\ldots,s_{M-1}\right\}}
	\PDF{\rv{y}|\rv{s}}{y|s}
\end{align}
where
\begin{align}
	\PDF{\rv{y}|\rv{s}}{y|s}
	&=
		\begin{cases}
			\PDF{\rv{t}_\mathrm{min}}{y-s},	&	y> s, \\
			0,	&	y \leq s.
		\end{cases}
\end{align}
\begin{proposition}[Density Function of First Arrival $H$-Noise]	\label{pro:pdf:fa}
Let $\rv{t}_{1}, \rv{t}_2, \ldots, \rv{t}_{N}$ be the i.i.d. $H$-noise. Then, the PDF of first arrival  $H$-noise $\PDF{\rv{t}_\mathrm{min}}{t}$ is given by
\begin{align}
	\PDF{\rv{t}_\mathrm{min}}{t}
	&=
		N
		\Fox{\pN_1+\pN_2}{\pM_1+\pM_2}{\pQ_1+\pQ_2}{\pP_1+\pP_2}
		{
		t;\pSeq_\rv{t}
		\Ket{\frac{^{}}{_{}}\left(a \pC_{\left(\omega_1\right)}\right)^{1/\left(\omega_1\omega_2\right)}}
		}
	\nonumber \\
	&\hspace{0.5cm}\times
		\left[
		\Fox{\pN_1+\pN_2+1}{\pM_1+\pM_2}{\pQ_1+\pQ_2+1}{\pP_1+\pP_2+1}
		{\frac{t}{\left(a\pC_{\left(\omega_1\right)}\right)^{1/\left(\omega_1\omega_2\right)}};\canOP{\pSeqcdf^{-1}}
			{
				\Bra{1}
				\pSeq_\rv{t}
			}
		}
		\right]^{N-1}.
\end{align}
\begin{proof}
The CDF $\CDF{\rv{t}_\mathrm{min}}{t}$ can be obtained as
\begin{align}	\label{eq:cdf:minHn}
	\CDF{\rv{t}_\mathrm{min}}{t}
	&=
		\Prob{\min\left\{\rv{t}_{1}, \rv{t}_2, \ldots,\rv{t}_{N}\right\}<t}
%	\nonumber \\
%	&=
%		1-\prod_{i=1}^N \Prob{\rv{t}_{\mathrm{Hn},i} > t}
%	\nonumber \\
%	&=
%		1-\prod_{i=1}^N \left(1-DF{\rv{t}_{\mathrm{Hn},i}}{t}\right)
	\nonumber \\
	&
=
		1-\left(1-\CDF{\rv{t}}{t}\right)^N.
\end{align}
Then, we have
\begin{align}
	\PDF{\rv{t}_\mathrm{min}}{t}
	&=
		d\CDF{\rv{t}_\mathrm{min}}{t}/dt
	\nonumber \\
&
	= N \PDF{\rv{t}}{t}\left(1-\CDF{\rv{t}}{t}\right)^{N-1}. 
\end{align}
from which and with the $H$-function representation, we complete the proof.
\end{proof}
\end{proposition}

Since $\PDF{\rv{t}_\mathrm{min}}{t}$ consists of the $H$-functions, the evaluation of exact error probability requires numerical calculation of an optimal detection threshold. Hence, we consider a simple upper bound expression, which can be achieved using a fixed detection threshold \cite{CTJS:15:CL,TJSW:18:COM}. %For ease of exposition, we provide the error probability analysis for the SHD, but a same procedure can be used to extend the analysis to $H$-diffusion.

\begin{theorem}[Upper bound on the SEP]	\label{thm:SEP:Hdiffusion}
The symbol error probability (SEP) $P_\mathrm{e}$ for $M$-ary and $N$-molecule transmission with the first arrival detection in $H$-diffusion is bounded as
\begin{align}	\label{eq:sep:Hdiffuion}
P_{\mathrm{e}}
&\leq
	\frac{M-1}{M}	
	\left[
		\Fox{\pN_1+\pN_2+1}{\pM_1+\pM_2}{\pQ_1+\pQ_2+1}{\pP_1+\pP_2+1}
		{\frac{T_{\mathrm{s}}/M}{\left(a\pC_{{\left(\omega_1\right)}}\right)^{1/\left(\omega_1\omega_2\right)}};	
			\pSeq_\mathrm{e}
		}
	\right]^N
\end{align}
where the parameter sequence %$\pSeq_\mathrm{e}$ is given by
%\begin{align}
%\pSeq_\mathrm{e}=
%\canOP{\pSeqcdf^{-1}}
%			{
%				\Bra{1}
%			\pSeq_\rv{t}
%			}.
%\end{align}
%
$\pSeq_\mathrm{e}=
\canOP{\pSeqcdf^{-1}}
			{
				\Bra{1}
			\pSeq_\rv{t}
			}$. 
\begin{proof}
For equally-likely symbols $s_i$, that is $\Prob{\rv{s}=s_i}=1/M$, we have
\begin{align}
	P_\mathrm{e}
	&=
		\frac{1}{M}
		\sum_{i=0}^{M-1}
		\Prob{\left.\hat{\rv{s}} \neq s_i \right|\rv{s}=s_i}
	\nonumber \\
	&
	\leq
		\frac{1}{M}
		\sum_{i=0}^{M-2}
		\Prob{\left.\rv{y}>\left(i+1\right)\frac{T_\mathrm{s}}{M}\right|\rv{s}=i\frac{T_\mathrm{s}}{M}}
	\nonumber \\
	&=
		\frac{M-1}{M}
		\left(
		1-
		\CDF{\rv{t}_\mathrm{min}}{\frac{T_\mathrm{s}}{M}}
		\right).
\end{align}
Using the CDF $\CDF{\rv{t}_\mathrm{min}}{t}$, which can be obtained from \eqref{eq:cdf:Hn} and \eqref{eq:cdf:minHn} in Proposition~\ref{pro:pdf:fa} as 
\begin{align}
\CDF{\rv{t}_\mathrm{min}}{t}
&=
	1-\left[
	\Fox{\pN_1+\pN_2+1}{\pM_1+\pM_2}{\pQ_1+\pQ_2+1}{\pP_1+\pP_2+1}
		{\frac{t}{\left(a\pC_{\left(\omega_1\right)}\right)^{1/\left(\omega_1\omega_2\right)}};\canOP{\pSeqcdf^{-1}}
			{
				\Bra{1}
				\pSeq_\rv{t}
			}
		}
	\right]^N,
\end{align}
we arrived at the desired result.
\end{proof}
\end{theorem}

%%%%%%%%%%%%%%%%%%%%%%%%%%%%%%%%%%%%%%%%%%%%%%%%%

\begin{definition}[Signal-to-Noise Power Ratio]		\label{def:snr}
The SNR for a molecular communication link can be defined as
\begin{align}
\mcsnr
\triangleq
\frac{1}{2\mathcal{G}}
\left(
\frac{T_\mathrm{s}}{\GMP{\rv{t}}}
\right)^2
\end{align}
where the constant $2\mathcal{G}$ is used to normalize the SNR corresponding to the standard SNR with Gaussian noise \cite[Table 1]{GPA:06:SP}.
\end{definition}

\subsubsection{Standard $H$-Diffusion}
For SHD, the SEP \eqref{eq:sep:Hdiffuion} is reduced in terms of $\mcsnr$ defined in Definition~\ref{def:snr} as
\begin{align}	\label{eq:sep:std}
P_\mathrm{e}
&\leq
	\frac{M-1}{M}
	\left[
		\Fox{2}{2}{4}{4}
		{
			\frac{M^2}{2\mathcal{G}^\star \mcsnr}
			;
			\hat{\pSeq}_\mathrm{e}
		}
	\right]^N
\end{align}
where 
\begin{align}
\hat{\pSeq}_\mathrm{e}
&=
	\left(
		\tfrac{4}{\alpha_1},
		1,
		\B{1}_4,
		\left(\B{1}_3,0\right),
		\left(2,\tfrac{2}{\alpha_1\omega_1\omega_2},\tfrac{1}{\omega_1\omega_2},\tfrac{2\alpha_2}{\omega_2}\right),
		\left(\tfrac{2}{\omega_2},\tfrac{2}{\omega_1\omega_2},\tfrac{1}{\omega_1\omega_2},2\right)
	\right)
\end{align}
with
$
\mathcal{G}^{\star}
	= 
		\mathcal{G}^{2\left(1-1/\alpha_1+\left(1-\alpha_2\right)\omega_1\right)/\left(\omega_1\omega_2\right)+1}$.

\subsubsection{High-SNR Expansions}
In the high-SNR regime, the SEP behaves as
\begin{align}
	P_\mathrm{e}
	&=
		\left(p_\infty \cdot \mcsnr\right)^{-s_\infty}+o\left(\mcsnr^{-s_\infty}\right),
		\quad\quad
		\mcsnr \rightarrow \infty
\end{align}
where the quantity $s_\infty$ denotes the \emph{high-SNR slope} of the SEP-$\mcsnr$ curve in a log-log scale, and $p_\infty$ represents the \emph{high-SNR power offset} in decibels of the SEP-$\mcsnr$ curve relative to a reference of $\mcsnr^{-s_\infty}$. 
\begin{corollary}[High-SNR Expansions]	\label{cor:highsnr}
At the high-SNR regime, two quantities $p_\infty$ and $s_\infty$ for SHD are given by
\begin{align}	
	s_\infty
	&=
		N\cdot \min\left\{\frac{\omega_2}{2},\frac{\omega_1\omega_2}{2}\right\}
	\label{eq:sinfty}
	\\
	p_\infty
	&=
		\left(\frac{1}{g\left(M,N\right)}\right)^{1/s_\infty}
	\label{eq:pinfty}
\end{align}
where
\begin{align}
g\left(M,N\right)
&=
	\begin{cases}
	\left(M-1\right)M^{N\omega_1\omega_2-1}
	\left(
		\frac{\GF{1-\omega_1}\GF{1/\alpha_1}}
		{\alpha_1\pi\GF{1-\alpha_2\omega_1}}
		\frac{\left(\mathcal{G}^{\star}\right)^{-\omega_1\omega_2/2}}{2^{\omega_1\omega_2/2-1}}
	\right)^N 
	& \omega_1 < 1, 
	\\
	\left(M-1\right)M^{N\omega_2-1}
		\left(
		\frac{
			\sin\left(\pi/\left(2\omega_1\right)\right)\GF{1-1/\omega_1}\GF{1/\left(\alpha_1\omega_1\right)}
			}
			{\alpha_1\pi\GF{1-\alpha_2}}
		\frac{\left(\mathcal{G}^{\star}\right)^{-\omega_2/2}}{2^{\omega_2/2-1}}
	\right)^N, &   \omega_1 > 1.
	\end{cases}
\end{align}
\begin{proof}
Using the algebraic asymptotic expansion of the $H$-function near zero, we have
\begin{align}	\label{eq:expansion:std}
		&
		\Fox{2}{2}{4}{4}
		{
			\frac{M^2}{2\mathcal{G}^\star \mcsnr}
			;
			\hat{\pSeq}_\mathrm{e}
		}
=
		\frac{4}{\alpha_1}
		\sigma^\star\left(\hat{\pSeq}_\mathrm{e}\right)
		\left(\frac{M^2}{2\mathcal{G}^\star \mcsnr}\right)^{\omega^\star\left(\hat{\pSeq}_\mathrm{e}\right)}
		+o\left(\mcsnr^{-\omega^\star\left(\hat{\pSeq}_\mathrm{e}\right)}\right),
		\quad\quad
		\mcsnr \rightarrow \infty 
\end{align}
where
\begin{align}
	\omega^\star\left(\hat{\pSeq}_\mathrm{e}\right)
	&=
		\min
		\left\{
			\frac{\omega_1 \omega_2}{2},
			\frac{\omega_2}{2}
		\right\}
	\\
	\sigma^\star\left(\hat{\pSeq}_\mathrm{e}\right)
	&=
		\begin{cases}
		\frac{\GF{1-\omega_1}\GF{1/\alpha_1}}{2\pi\GF{1-\alpha_2\omega_1}},	&	\omega_1<1 \\
		\frac{\sin\left(\pi/\left(2\omega_1\right)\right)\GF{1-1/\omega_1}\GF{1/\left(\alpha_1\omega_1\right)}}{2\pi\GF{1-\alpha_2}}, 	&	\omega_1> 1.
		\end{cases}
\end{align}
From \eqref{eq:expansion:std} and some manipulations, the desired result is obtained.
\end{proof}
\end{corollary}

%%%%%%%%%%%%%%%%%%%%%%%%%%%%%%%%%%%%%%%%%%%%%%%%%

\iffalse
The high-SNR expansions of the SEP $P_\mathrm{e}$ for typical anomalous diffusion in Table~\ref{table:TDM:HD} are tabulated in Table~\ref{table:expansions} as special cases of \eqref{eq:sinfty} and \eqref{eq:pinfty}.
\fi

\begin{remark}[High-SNR Slope]
As shown in Corollary~\ref{cor:highsnr}, the high-SNR slope $s_\infty$  increases linearly with the number of released molecules $N$. This can be interpreted as the benefits arising from consuming molecular resources, which is synonymous with \emph{transmit diversity} in wireless multiple-antenna systems.
\end{remark}

 \begin{figure}[t!]
    \centerline{\includegraphics[width=0.55\textwidth]{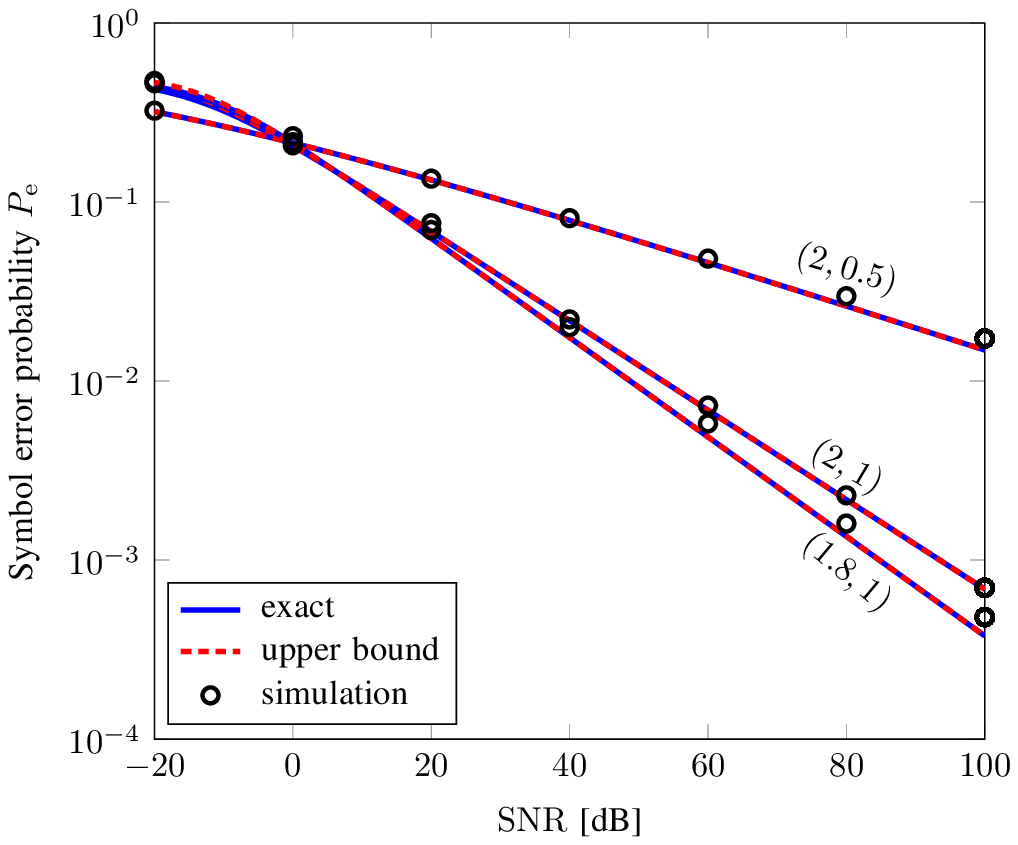}}
    %\vspace{-0.3cm}
\caption{
       SEP $P_\mathrm{e}$ as a function of $\mcsnr$ [dB] in the $\left(\alpha_1,\alpha_2\right)$-SHD for:  
       i) $\left(\alpha_1,\alpha_2\right)=\left(2,1\right)$; 
       ii) $\left(\alpha_1,\alpha_2\right)=\left(2,0.5\right)$; and  
       iii) $\left(\alpha_1,\alpha_2\right)=\left(1.8,1\right)$; 
       with $N=1$ and $M=2$.
}
\label{fig:10}
\end{figure}

 \begin{figure}[t!]
    \centerline{\includegraphics[width=0.55\textwidth]{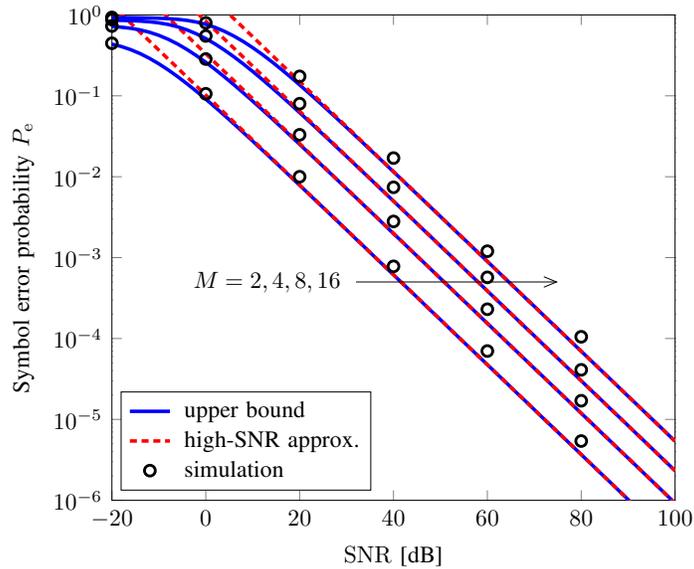}}
    %\vspace{-0.3cm}
\caption{
       SEP $P_\mathrm{e}$ as a function of $\mcsnr$ [dB] in the $\left(1.8,1\right)$-SHD for $M=2$, $4$, $8$, $16$, with $N=2$.
       }
\label{fig:11}
\end{figure}

\subsection{Numerical Examples}
Fig.~\ref{fig:10} shows the SEP as a function of $\mcsnr$ [dB] in $\left(\alpha_1,\alpha_2\right)$-SHD for the three diffusion scenarios with single molecule transmission ($N=1$) and binary modulation ($M=2$). Obviously, the SEP decreases with SNR, which implies that we can improve the molecular communication reliability with a large symbol time $T_\mathrm{s}$ (with a low  symbol rate). We can observe that the high-SNR slope in $\left(\alpha_1,\alpha_2\right)$-SHD can be determined by  $s_\infty=\alpha_2/\left(2\alpha_1\right)$ for $\alpha_1>1$, as stated in Corollary~\ref{cor:highsnr}. In this example, $s_\infty=0.25$, $0.125$, and $0.278$ for $\left(2,1\right)$-, $\left(2,0.5\right)$-, and $\left(1.8,1\right)$-SHD, respectively. Therefore, the SEP in superdiffusion outperforms that in other scenarios in the high-SNR regime due to the large value of $s_\infty$. It is also noteworthy that the upper bound becomes tight at the high-SNR regime since the detection thresholds approach to $i T_s/M$, $i=1,\ldots, M-1$ \cite{TJSW:18:COM}.

 \begin{figure}[t!]
    \centerline{\includegraphics[width=0.55\textwidth]{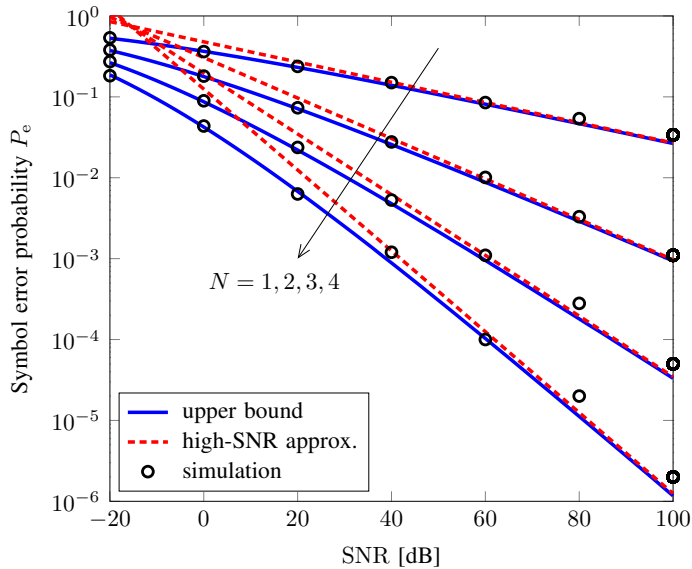}}
    %\vspace{-0.3cm}
\caption{
       SEP $P_\mathrm{e}$ as a function of $\mcsnr$ [dB] in the $\left(2,0.5\right)$-SHD for $N=1$, $2$, $3$, $4$, with $M=4$.
}
\label{fig:12}
\end{figure}

\begin{figure}[t!]
    \centerline{\includegraphics[width=0.55\textwidth]{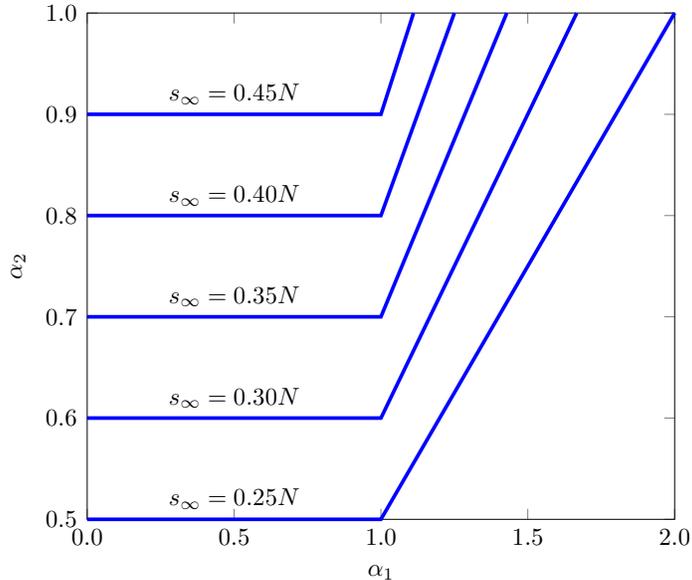}}
    %\vspace{-0.3cm}
\caption{
       High-SNR slope $s_\infty$ of SEP $P_\mathrm{e}$ as a function of $\left(\alpha_1,\alpha_2\right)$ in the $\left(\alpha_1,\alpha_2\right)$-SHD. $\left(\alpha_1,\alpha_2\right)$-line is for $s_\infty=0.45N$, $0.40N$, $0.35N$, $0.30N$, and $0.25N$.
       }
\label{fig:13}
\end{figure}

Fig.~\ref{fig:11} shows the SEP $P_\mathrm{e}$ as a function of $\mcsnr$ [dB] in the $\left(1.8,1\right)$-SHD with $N=2$ for $M=2$, $4$, $8$, $16$. The figure shows that we can boost the data rate with modulation order $M$ by sacrificing the reliability of SEP performance. Fig.~\ref{fig:12} demonstrates the effect of the number of released molecules on error performance, where the SEP $P_\mathrm{e}$ is a  function of $\mcsnr$ [dB] in the $\left(2,0.5\right)$-SHD with $M=4$ for $N=1$, $2$, $3$, $4$. The  released molecule diversity can be obtained linearly with multiple released molecules. In this example, the high-SNR slopes are equal to $s_\infty=0.125$, $0.25$, $0.375$, and $0.5$ for $N=1$, $2$, $3$, and  $4$, respectively. As can be seen from both Figs.~\ref{fig:11} and \ref{fig:12}, the derived high-SNR expansion expression has close agreement in slope and power offset of the SEP curves. The high-SNR slope $s_\infty$ of SEP in the $\left(\alpha_1,\alpha_2\right)$-SHD can be further examined by referring to Fig.~\ref{fig:13}, where $\left(\alpha_1,\alpha_2\right)$-lines are depicted for $s_\infty=0.45N$, $0.40N$, $0.35N$, $0.30N$, and $0.25N$. The high-SNR slope $s_\infty$ increases with $\alpha_2$, while it decreases with $\alpha_1$ in the region $\alpha_1 >1$, as expected.

%%%%%%%%%%%%%%%%%%%%%%%%%%%%%%%%%%%%%%%%%%

\section{Conclusion}
In this paper, we developed a new mathematical framework for modeling and analysis in molecular communication with anomalous diffusion. We first systemized the method to generate the PDF---involved in the subordination law---of the molecule's position evolving in time by introducing the general class of diffusion processes, namely $H$-diffusion. The $H$-diffusion modeling subsumes most known anomalous diffusion models obtained from two $H$-variates, which play a role in explaining the anomalous evolution of molecules in space and in time, respectively. We then provided the new notion of molecular noise, namely $H$-noise, to account for the statistical properties of uncertainty of the random propagation time under $H$-diffusion laws. The error performance in the subdiffusion scenario outperforms that in the low-SNR regime, while the superdiffusion scenario is outperformed in the high-SNR regime.
In summary, our framework proved by $H$-theory serves a method to model a molecular communication channel with anomalous diffusion and unified analysis for emerging nanoscale communication.

This work opens several open problems. For a practice, the multi-dimensional anomalous diffusion channel model should be considered with suitable initial and boundary conditions depending on various molecular communication system setups. Since the molecules obey different diffusion rules compared to normal diffusion, new inter-symbol-interference avoidance schemes, synchronization schemes between nanomahcine, and receiver designs are required.

\newpage

\appendices

\section{Glossary of Notations and Symbols}	\label{sec:appendix:NS}

\begin{basedescript}{\desclabelwidth{2.8cm}}
{

\item[~$\R$]

Real numbers \\[-0.8cm]

\item[~$\R_+$]

Nonnegative real numbers \\[-0.8cm]

\item[~$\R_{++}$]

Positive real numbers \\[-0.8cm]

%\item[$\Stable{\Salpha}{\Sbeta}{\Sgamma}{\Smu}$]

%Stable distribution with characteristic exponent $\alpha \in \left(0,2\right]$, 
%skewness parameter $\beta \in \left[-1,1\right]$, 
%scale parameter $\gamma \in \left[0,\infty\right)$,
%and location parameter $\mu \in \R$

\item[~$\B{1}_n$] 

All-one sequence or vector of $n$ elements \\[-0.8cm]

\item[~$\B{0}_n$] 

All-zero sequence or vector of $n$ elements \\[-0.8cm]

\item[~$o\left(\cdot\right)$]

Bachmann--Landau notation: \\[-0.8cm]
\begin{flalign}
&f\left(x\right)
=o\left(g\left(x\right)\right)
	\quad
	\left(x \rightarrow x_0\right)
%\nonumber \\
\quad
    \Leftrightarrow ~
    \lim_{x\rightarrow x_0}\frac{f\left(x\right)}{g\left(x\right)}=0
&
\end{flalign}
~\\[-1.8cm]
\item[~$\deq$]

Distributional equality \\[-0.8cm]

\item[~$\doteq$]

Asymptotically exponential equality \\[-0.8cm]
\begin{flalign}
&f\left(x\right) \doteq x^y
    ~\Leftrightarrow~
    \lim_{x \rightarrow \infty}
        \frac{\log f\left(x\right)}{\log x}
    =y
&
\end{flalign}
~\\[-0.8cm]
where $y$ is the exponential order of $f\left(x\right)$.\\[-0.8cm]

\item[~$\EX{\cdot}$] 

Expectation operator \\[-0.8cm]

%\item[~$\Var{\cdot}$] 

%Variance operator \\[-0.8cm]

\item[~$\PDF{\rv{x}}{x}$]

Probability density function of $\rv{x}$ \\[-0.8cm]

\item[~$\CDF{\rv{x}}{x}$] 

Cumulative distribution function of $\rv{x}$  \\[-0.8cm]

\item[~$\CF{\rv{x}}{x}$]

Characteristic function of $\rv{x}$ \\[-0.8cm]

\item[~$\delta\left(x\right)$]

Dirac delta function \\[-0.8cm]

%\item[~$Q\left(\cdot\right)$] 

%$Q$-function  \\[-0.8cm]

\item[~$\GF{\cdot}$] 

Gamma function \cite[eq.~(8.310.1)]{GR:07:Book} \\[-0.8cm]

%\item[~$I_x\left(a,b\right)$] 

%Regularized incomplete beta function \cite[eq.~(8.392)]{GR:07:Book}  \\[-0.8cm]

%\item[~$\GMLF{\alpha}{\beta}{t}$]

%Generalized Mittag-Leffler function \cite{GKMR:14:Book}  \\[-0.8cm]

%\item[~$E_{\nu}\left(t\right)$]

%Mittag-Leffler function \cite{GKMR:14:Book}  \\[-0.8cm]

%\item[~$W_{\lambda,\mu}\left(t\right)$]

%Wright function of the second kind \cite{MMP:10:JDE}  \\[-0.8cm]

%\item[~$M_\nu\left(t\right)$]

%$M$-Wright function \cite{MMP:10:JDE}  \\[-0.8cm]

\item[~$H^{\pM,\pN}_{\pP,\pQ}{\left[\cdot\right]}$]

Fox's $H$-function \cite{JSW:15:IT}: \\[-0.7cm]
%For notational simplicity, we shall use the notation
\begin{flalign}
\Fox{\pM}{\pN}{\pP}{\pQ}{x;\pSeq}
&=
	\pK
	\FoxH{\pM}{\pN}{\pP}{\pQ}{\pC x}{
        \left(\pA{1},\pS{1}\right),
		\left(\pA{2},\pS{2}\right),
		\ldots,
		\left(\pA{\pP},\pS{\pP}\right)}{
		\left(\pB{1},\pT{1}\right),
		\left(\pB{2},\pT{2}\right),
		\ldots,
		\left(\pB{\pQ},\pT{\pQ}\right)}
	\nonumber\\
&=
	\pK
	\FoxH{\pM}{\pN}{\pP}{\pQ}{\pC x}
		{\left(\aV,\sV\right)}
		{\left(\bV,\tV\right)}
&
\end{flalign}
where the parameter sequence is 
$
\pSeq
=
	\left(\pK,\pC,\aV,\bV,\sV,\tV\right)
$
with\\[-0.7cm]
\begin{flalign}
&
\begin{cases}
\aV=\left(\pA{1},\pA{2},\ldots,\pA{\pN},\pA{\pN+1},\pA{\pN+2}\ldots,\pA{\pP}\right) 
\\[-0.3cm]
\bV=\left(\pB{1},\pB{2},\ldots,\pB{\pM},\pB{\pM+1},\pB{\pM+2},\ldots,\pB{\pQ}\right)
\\[-0.3cm]
\sV=\left(\pS{1},\pS{2},\ldots,\pS{\pN},\pS{\pN+1},\pS{\pN+2},\ldots,\pS{\pP}\right)
\\[-0.3cm]
\tV=\left(\pT{1},\pT{2},\ldots,\pT{\pM},\pT{\pM+1},\pT{\pM+2},\ldots,\pT{\pQ}\right)
\end{cases}
&
\end{flalign}
A Mellin-Barnes type integral form of Fox's $H$-function is\\[-0.8cm]
\begin{flalign}
\Fox{\pM}{\pN}{\pP}{\pQ}{x;\pSeq}
&=
	\frac{1}{2\pi \jmath}
    \int_\mathfrak{L}
    \theta\left(s\right)
    x^s
    ds,\quad x\neq 0
&
\end{flalign}
where $\mathfrak{L}$ is a suitable contour, $\jmath=\sqrt{-1}$, 
$
x^s
=
    \exp\left\{s\left(\ln\left|x\right|+\jmath \arg x\right)\right\}
$,
and 
\begin{flalign}
\theta\left(s\right)
&=
    \frac{
            \prod_{j=1}^{\pM}\GF{\pB{j}-\pT{j}s}
            \prod_{j=1}^{\pN}\GF{1-\pA{j}+\pS{j}s}
         }
         {
            \prod_{j=\pM+1}^{\pQ}\GF{1-\pB{j}+\pT{j}s}
            \prod_{j=\pN+1}^{\pP}\GF{\pA{j}-\pS{j}s}
         }
&
\end{flalign}

\item[~$\FoxHT{\pM,\pN}{\pP,\pQ}{\pSeq}{f\left(t\right)}{s}$]

$H$-transform of a function $f\left(t\right)$ with Fox's $H$-kernel of the order sequence \\[-0.8cm]
\item[]
$\oSeq=\left(\pM, \pN, \pP, \pQ\right)$ and the parameter sequence $\pSeq=\left(\pK,\pC,\aV,\bV,\sV,\tV\right)$ \cite{JSW:15:IT}: \\[-0.7cm]
\begin{flalign} \label{eq:Def:HT}
\FoxHT{\pM,\pN}{\pP,\pQ}{\pSeq}{f\left(t\right)}{s}
&=
	\pK
    	\int_0^\infty
        	\FoxH{\pM}{\pN}{\pP}{\pQ}
        	{\pC st}
        	{\left(\aV, \sV\right)}
        	{\left(\bV, \tV\right)}
        	f\left(t\right)
        dt,
        \quad
        s>0
&
\end{flalign}

\item[~$\FoxV{\pM}{\pN}{\pP}{\pQ}{\pSeq}$]

$H$-variate with the order sequence $\oSeq=\left(\pM,\pN,\pP,\pQ\right)$ and the parameter sequence \\[-0.8cm]
\item[]
$\pSeq = \left(\pK, \pC, \aV, \bV, \sV, \tV \right)$ \cite{JSW:15:IT}: if $\rv{x} \sim \FoxV{\pM}{\pN}{\pP}{\pQ}{\pSeq}$, then \\[-0.7cm]
\begin{flalign} \label{eq:Def:FV}
\PDF{\rv{x}}{x}
&=
        \pK
        \FoxH{\pM}{\pN}{\pP}{\pQ}
        {\pC x}
        {\left(\aV,\sV\right)}
        {\left(\bV,\tV\right)},
        \qquad
        x \geq 0
&
\end{flalign} \\[-0.7cm]
with the set of parameters satisfying a distributional structure such that \\[-0.8cm]
\item[]
$\PDF{\rv{x}}{x} \geq 0$ for all $x \in \R_+$ and $\FoxHT{\pM,\pN}{\pP,\pQ}{\pSeq}{1}{1}=1$ \\[-0.8cm]
}
\end{basedescript}

\section{Special Functions}	\label{sec:appendix:SF}

In this appendix, we briefly introduce the special functions which are frequently used in the context of diffusion theory, fractional calculation theory, and molecular communication.

\subsection{$M$-Wright Function}	\label{sec:appendix:SF:A}
The Wright function (of the second kind) is defined as
\begin{align}
W_{\lambda,\mu}\left(t\right)
=
	\sum_{n=0}^\infty
	\frac{t^n}{n! \GF{\lambda n + \mu}}
\end{align}
where $\lambda > -1$, $\mu \in \C$, and $t \in \C$.
The $M$-Wright function $M_\nu\left(t\right)$ is the one of the auxiliary function of the Wright function by setting $\lambda=-\nu$ and $\mu=1-\nu$, whose series representation is given by
\begin{align}
M_\nu\left(t\right)
&=
	\sum_{n=0}^\infty
	\frac{\left(-t\right)^n}{n!\GF{-\nu n + \left(1-\nu\right)}}
	\nonumber \\
&
=
	\frac{1}{\pi}
	\sum_{n=1}^\infty
	\frac{\left(-t\right)^{n-1}}{\left(n-1\right)!}\GF{\nu n}\sin\left(\pi n \nu\right)
\end{align}
where $\nu$ is defined on the positive real axis for $0<\nu<1$. Note that $M_{\nu=1}\left(t\right)=\delta\left(z\right)$. It also can be represented in terms of the $H$-function as
\begin{align}
M_\nu\left(t\right)
=
	\Fox{1}{0}{1}{1}{t;\pSeq_\mathrm{MW}=\left(1,1,1-\nu,0,\nu,1\right)}.
\end{align}
An important particular case is for $\nu=1/2$ where the $M$-Wright function reduces to
\begin{align}
M_{\nu=1/2}\left(t\right)
&=
	\Fox{1}{0}{1}{1}{t;\left(1,1,1/2,0,1/2,1\right)}
\nonumber \\
&
=
	\frac{1}{\sqrt{\pi}}
	\,
	\exp\left(-t^2/4\right)
\end{align}
which can be interpreted as a natural generalization of the Gaussian density for fractional diffusion processes. Since the $M$-Wright function has a relation to Mittag-Leffler function $E_\nu\left(z\right)$ by means of the Laplace transform 
\begin{align}
\LT{M_{\nu}\left(t\right)}{s}
&=
\FoxHT{1,0}{0,1}{\left(1,1,\text{--},0,\text{--},1\right)}{M_{\nu}\left(t\right)}{s}
\nonumber \\
&
=
\Fox{1}{1}{1}{2}{s;\left(1,1,0,\B{0}_2,1,\left(1,\nu\right)\right)}
\nonumber \\
&
=
\sum_{n=0}^{\infty}
\frac{\left(-s\right)^n}{\GF{\nu n +1}}
\nonumber \\
&
=
E_\nu\left(-s\right),
\end{align}
the $M$-Wright function for nonnegative variable $t$ is known as a non-Markovian model to generalize the evolution in time of fractional diffusion processes. The $M$-Wright function is also related to the inverse stable subordinator to the generalized grey Brownian motion \cite{Pag:13:FCAA, MMP:10:JDE,PS:14:CAIM}, and hence the $M$-Wright function for the symmetric random variable can be represented as all solutions of EK-FD process.

\subsection{Mittag-Leffler Function}		\label{sec:appendix:SF:B}

The Mittag-Leffler function appears as the solution of fractional differential equations and fractional order integral equations \cite{Lef:05:AM, Cah:13:CSTM, Gor:09:Proc, GKMR:14:Book, Lin:98:SPI}. 
The series and $H$-function representations for the generalized Mittag-Leffler function $E_{\alpha,\beta}\left(t\right)$ are given by
\begin{align}
E_{\alpha,\beta}\left(t\right)
&=
	\sum_{n=0}^\infty
	\frac{t^n}{\GF{\alpha n + \beta}}
\nonumber \\
&
=
	\Fox{1}{1}{1}{2}{t;\pSeq_\mathrm{GML}=\left(1,-1,0,\left(0,1-\beta\right),1,\left(1,\alpha\right)\right)}
\end{align}
where $\alpha,\beta \in \C$ and $\Re\left(\alpha\right), \Re\left(\beta\right) > 0$. When $\alpha=\nu$ and $\beta=1$, $E_{\alpha,\beta}\left(t\right)$ reduces to Mittag-Leffler function, denoted by $E_{\nu}\left(t\right)$, as
\begin{align}
E_{\nu}\left(t\right)
&=
	\sum_{n=0}^\infty
	\frac{t^n}{\GF{\nu n + 1}}
\nonumber \\
&
=
	\Fox{1}{1}{1}{2}{t;\pSeq_\mathrm{ML}=\left(1,-1,0,\B{0}_2,1,\left(1,\nu\right)\right)}.
\end{align}
Since $E_\nu\left(t\right)$ is a increasing function for all $\nu\in \left(0,1\right]$ with the convergences $E_\nu\left(-\infty\right)=0$ and $E_\nu\left(0\right)=1$, the cumulative distribution function of a probability measure on the nonnegative real numbers can be defined as $\CDF{\rv{t}}{t;\nu}=1-E_\nu\left(-t^\nu\right)$ called the Mittag-Leffler distribution of order $\nu$. For $\nu \in \left(0,1\right)$, the Mittag-Leffler distribution of order $\nu$ is a heavy-tailed generalization of the exponential distribution since the Laplace transform of 
\begin{align}	\label{eq:mld}
\rv{t}\sim\FoxV{1}{1}{1}{2}{\pSeq_\mathrm{MLD}=\left(\tfrac{1}{\nu},1,1-\tfrac{1}{\nu},\left(1-\tfrac{1}{\nu},0\right),\tfrac{1}{\nu},\left(\tfrac{1}{\nu},1\right)\right)}
%\rv{t}\sim\FoxV{2}{1}{2}{3}{\pSeq_\mathrm{MLD}=\left(\tfrac{1}{\nu},1,\left(-\tfrac{1}{\nu},-1\right),\left(0,-\tfrac{1}{\nu},-1\right),\left(\tfrac{1}{\nu},1\right),\left(1,\tfrac{1}{\nu},1\right)\right)}
\end{align}
has a form%\footnote{The PDF of Mittag-Leffler distribution in \eqref{eq:mld} can be obtained from the cumulative distribution function expression using the differentiation operator of the $H$-function \cite[Property~4]{JSW:15:IT} such that
%$$
%\rv{t}\sim\FoxV{2}{1}{2}{3}{\pSeq=\left(\tfrac{1}{\nu},1,\left(-\tfrac{1}{\nu},-1\right),\left(0,-\tfrac{1}{\nu},-1\right),\left(\tfrac{1}{\nu},1\right),\left(1,\tfrac{1}{\nu},1\right)\right)}.
%$$}
\begin{align}
\EX{e^{-\rv{t}s}}=\frac{1}{1+s^{\nu}}.
\end{align}
Note that $\rv{t}$ is a completely skewed geometric stable distribution and also an exponential distribution with the order $\nu=1$.

\section{$H$-Representation Theory of Strictly Stable Distributions}	\label{sec:appendix:Stable}

The stable distribution is a rich class of probability distributions that allow skewness and
heavier (algebraic) tails.  
This distribution has been used in modeling and analyzing physical, statistical, engineering, and economic systems. However, due to lack of analytical expression for the density function of stable distributions for all but a few cases---for example, normal, Cauchy, and L\'evy distributions---the use of stable distributions meets many technical challenges. In this appendix, we provide an analytical expression for the stable distributions in terms of $H$-functions.

Let $\rv{x}_1$ and $\rv{x}_2$ be independent copies of a random variable $\rv{x}$. Then $\rv{x}$ is said to be \emph{stable} if, for any constants $a>0$ and $b>0$,
\begin{align}	\label{eq:def:stable}
a\rv{x}_1 + b\rv{x}_2 \deq c\rv{x} + d
\end{align}
for constants $c>0$ and $d$. The distribution is said to be \emph{strictly stable} when \eqref{eq:def:stable} holds with $d=0$.
In what follows, we use $\Stable{\Salpha}{\Sbeta}{\Sgamma}{\Smu}$ to denote the distribution of a stable random variable with

    \vspace{0.5cm}
    \hspace{3cm}\begin{tabular}{ll}
    $\Salpha \in \left(0,2\right]$ & characteristic exponent (index of stability) \\
    $\Sbeta \in \left[-1,1\right]$ & skewness parameter \\
    $\Sgamma \in \left[0,\infty\right)$ & scale (dispersion) parameter \\
    $\Smu \in \R $ & location parameter. \\
    \end{tabular}
    \vspace{0.5cm}

\noindent
In general, the canonical characteristic function of stable distribution $\rv{x}\sim\Stable{\Salpha}{\Sbeta}{\Sgamma}{\Smu}$ is
\begin{align}
        \CF{\rv{x}}{\omega}
        &=
        \begin{cases}
        \exp\left\{-\Sgamma\left|\omega\right|^\Salpha
        \left[1-\jmath\Sbeta\sign\left(\omega\right)\tan\left(\frac{\pi\Salpha}{2}\right)
        \right]
        +\jmath\Smu\omega\right\},    &   \textrm{if}~\Salpha \neq 1 \\
        \exp\left\{-\Sgamma\left|\omega\right|
        \left[1+\jmath\Sbeta\sign\left(\omega\right)\frac{2}{\pi}\ln|\omega|\right]
        +\jmath\Smu\omega\right\},    &   \textrm{if}~\Salpha = 1.
        \end{cases}
\end{align}

We discuss the special cases for the stable distribution as follows:
\begin{itemize}

\item   

When $\Salpha \neq 1$: The representation problem of stable densities with $\Salpha\neq 1$ is equivalent to finding the inverse Fourier transform of the form
    \begin{align}
        \psi_{\Salpha,\theta}\left(\omega\right)
        =
            \exp\left\{-\left|\omega\right|^\Salpha e^{-\jmath \frac{\pi}{2}\theta \sign\left(\omega\right)}\right\}
    \end{align}
    where $\left|\theta\right|\leq\min\left\{\Salpha,2-\Salpha\right\}$. The PDF of $\rv{x}\sim\Stable{\Salpha}{\Sbeta}{\Sgamma}{\Smu}$ for $\Salpha \neq 1$ is given by
    \begin{align}
        \PDF{\rv{x}}{x}
        =
            \Fox{1}{1}{2}{2}{\left|x-\Smu\right|; \pSeq_{\Salpha,\Sbeta,\Sgamma,\Smu}}
    \end{align}
    where the parameter sequence $\pSeq_{\Salpha,\Sbeta,\Sgamma,\Smu}$ is
    \begin{align}		\label{eq:pSeq:stable}
    \pSeq_{\Salpha,\Sbeta,\Sgamma,\Smu}
    &=
        \biggl(
                \frac{\omega_{\Salpha,\Sbeta,\Sgamma}}{\Salpha},
                \omega_{\Salpha,\Sbeta,\Sgamma},
                \aV_\mathrm{s}, \bV_\mathrm{s},\sV_\mathrm{s}, \tV_\mathrm{s}
        \biggr)
    \end{align}
    with
    \begin{align}
    \begin{cases}
    \aV_\mathrm{s}
    &\hspace{-0.35cm}=
    		\Bigl(1-\dfrac{1}{\Salpha},\dfrac{1}{2}-\sign\left(x-\Smu\right)\theta_{\Salpha,\Sbeta}\Bigr)\\
    \bV_\mathrm{s}
    &\hspace{-0.35cm}=
    		\Bigl(0,\dfrac{1}{2}-\sign\left(x-\Smu\right)\theta_{\Salpha,\Sbeta}\Bigr)\\
    \sV_\mathrm{s}
    &\hspace{-0.35cm}=
                \Bigl(\dfrac{1}{\Salpha},\dfrac{1}{2}+\sign\left(x-\Smu\right)\theta_{\Salpha,\Sbeta}\Bigr)\\
    \tV_\mathrm{s}
    &\hspace{-0.35cm}=
                \Bigl(1,\dfrac{1}{2}+\sign\left(x-\Smu\right)\theta_{\Salpha,\Sbeta}\Bigr)
    \end{cases}
    \end{align}
and
    \begin{align}
    \omega_{\Salpha,\Sbeta,\Sgamma}
    &=
        \left(\Sgamma\sqrt{1+\Sbeta^2\tan^2\left(\pi\Salpha/2\right)}\right)^{-1/\Salpha}
    \\
    \theta_{\Salpha,\Sbeta}
    &=
        \frac{1}{\pi\Salpha}
        \tan^{-1}\left(\Sbeta\tan\frac{\pi\Salpha}{2}\right).
    \end{align}
The result shows that \emph{all} stable densities with $\Salpha\neq 1$ can be expressed in terms of $H$-functions.

\item		

When $\Salpha=1$ (\emph{Terra Incognito}): Due to the presence of the logarithm $\ln\left|\omega\right|$ in the characteristic function, little information is available about the distributional structure for this case. If $\rv{x}\sim \Stable{1}{\Sbeta}{\Sgamma}{\Smu}$ is \emph{strictly} stable (i.e., $\Sbeta=0$), we have
    \begin{align}
    \PDF{\rv{x}}{x}
    =
        \Fox{1}{1}{2}{2}{\left|x-\Smu\right|;\pSeq_{1,0,\Sgamma,\Smu}}
    \end{align}
    where $\pSeq_{1,0,\Sgamma,\Smu}$ reduces to
    \begin{align}
    \pSeq_{\Sgamma}
    =
        \left(
                \frac{1}{\Sgamma},\frac{1}{\Sgamma},\left(0,\frac{1}{2}\right),\left(0,\frac{1}{2}\right),\left(1,\frac{1}{2}\right),\left(1,\frac{1}{2}\right)
        \right).
    \end{align}
This result shows that \emph{only strictly} stable densities with $\Salpha= 1$ are known in terms of $H$-functions.

\item

Nonnegative Stable Laws: Since all the one-sided stable distributions are \emph{extremal} (i.e., $\Sbeta=\pm 1$ with the support $\R_\pm$), $\rv{x}\sim\Stable{\Salpha}{\Sbeta}{\Sgamma}{\Smu}$ is nonnegative if and only if $0<\Salpha<1$, $\Sbeta=1$, and $\Smu=0$ (necessarily strictly stable). Hence,
    \begin{align}
    \PDF{\rv{x}}{x}
    =
        \Fox{0}{1}{1}{1}{x;\pSeq_\mathrm{pdf}^\star},   \quad x\geq0
    \end{align}
    where
    \begin{align}
    \pSeq_\mathrm{pdf}^\star
    =
        \left(
                \frac{\omega_{\Salpha,1,\Sgamma}}{\Salpha},
                \omega_{\Salpha,1,\Sgamma},
                1-\frac{1}{\Salpha},0,\frac{1}{\Salpha},1
        \right).
    \end{align}
    Using the language of $H$-functions, the characteristic function of $\rv{x}$ can be expressed again in terms of $H$-function as follows:
    \begin{align}
    \CF{\rv{x}}{\omega}
    =
        \Fox{1}{1}{1}{2}{-\jmath\omega;\pSeq_\mathrm{cf}^\star}
    \end{align}
    where
    \begin{align}
    \pSeq_\mathrm{cf}^\star
    =
        \left(
                \frac{1}{\Salpha},\frac{1}{\omega_{\Salpha,1,\Sgamma}},0,\B{0}_2,1,\left(\frac{1}{\Salpha},1\right)
        \right).
    \end{align}
\emph{All nonnegative} stable random variables are $H$-variates. The aggregate interference power in a Poisson field belongs to this class of stable laws.
\end{itemize}


\end{document}